\documentclass{jfm}

\usepackage[dvips]{epsfig}

\usepackage{mathrsfs}
\usepackage{amsmath}
\usepackage{amssymb}
\usepackage{multicol}
\usepackage{float}
\usepackage{makeidx}
\usepackage{layout}
\usepackage{array}
\usepackage{boxedminipage}
\usepackage{latexsym}
\usepackage{indent first}

\ifCUPmtlplainloaded \else
  \checkfont{eurm10}
  \iffontfound
    \IfFileExists{upmath.sty}
      {\typeout{^^JFound AMS Euler Roman fonts on the system,
                   using the 'upmath' package.^^J}%
       \usepackage{upmath}}
      {\typeout{^^JFound AMS Euler Roman fonts on the system, but you
                   dont seem to have the}%
       \typeout{'upmath' package installed. JFM.cls can take advantage
                 of these fonts,^^Jif you use 'upmath' package.^^J}%
      }
  \else
  \fi
\fi

\ifCUPmtlplainloaded \else
  \checkfont{msam10}
  \iffontfound
    \IfFileExists{amssymb.sty}
      {\typeout{^^JFound AMS Symbol fonts on the system, using the
                'amssymb' package.^^J}%
       \usepackage{amssymb}%

      }{}
  \fi
\fi

\ifCUPmtlplainloaded \else
  \IfFileExists{amsbsy.sty}
    {\typeout{^^JFound the 'amsbsy' package on the system, using it.^^J}%
     \usepackage{amsbsy}}
    {}
\fi

\newcommand\Rey{\mbox{\textit{Re}}}  

\newsavebox{\astrutbox}
\sbox{\astrutbox}{\rule[-5pt]{0pt}{20pt}}

\newcommand\p{\ensuremath{\partial}}

\title[Gravity-Capillary Solitary Waves]
{Dynamics of Gravity-Capillary Solitary Waves in Deep Water}
\author[Z. Wang and P. A. Milewski]
{Z\ls H\ls A\ls N\ns W\ls A\ls N\ls G$^1$ \and P\ls A\ls U\ls L\ns A.\ns
M\ls I\ls L\ls E\ls W\ls S\ls K\ls I$^2$}

\affiliation{$^1$Department of Mathematics, University of Wisconsin-Madison, Madison WI, 53706, USA\\[\affilskip]
$^2$Department of Mathematical Sciences, University of Bath, Bath, BA2 7AY, U.K.}
\date{}

\begin{document}

\maketitle

\begin{abstract}

The dynamics of solitary gravity-capillary
water waves propagating on the surface of a three-dimensional fluid
domain is studied numerically. In order to accurately compute complex time dependent
solutions, we simplify the full potential flow problem by taking a
cubic truncation of the scaled Dirichlet-to-Neumann operator for the
normal velocity on the free surface. This approximation agrees
remarkably well with the full equations for the bifurcation curves, wave profiles
and the dynamics of solitary waves for a two-dimensional fluid domain. Fully
localised solitary waves are then computed in the three-dimensional problem and the stability and interaction
of both line and localized solitary waves are investigated via numerical time integration of the equations. The
solitary wave branches are indexed by their finite energy at small amplitude, and
the dynamics of the solitary waves is complex involving nonlinear focussing of wave
packets, quasi-elastic collisions, and the generation of propagating, spatially localised, time-periodic structures (breathers).

\end{abstract}

\section{Introduction}

Deep water gravity-capillary waves are relevant in a range of applications,
including the understanding of the generation of waves
by wind and the interpretation of satellite remote sensing data
(see, for example {\cite{ZX}} and {\cite{WJW}}). The study of solitary waves in this regime is of
particular theoretical interest since it is \textit{only} under the joint
effects of surface tension and gravity that localised water waves in three-dimensions have been found
in the water wave problem. Such localised waves have recently been observed in the
experiments of \cite{CDAD1}. In the deep water limit, these Gravity-Capillary (GC)
``wavepacket" solitary waves are of a fundamentally different nature
from ``long" solitary waves which are described approximately by the
Korteweg--de Vries (KdV), Kadomtsev-Petviashvilli (KP) and related
equations in shallow water. The later bifurcate from linear waves at
zero wavenumber whereas the former bifurcate at a finite wavenumber,
and hence their oscillatory nature. These oscillatory solitary waves
were seen by \cite{AS}  and \cite{A} to be approximately described by solitary wave
solutions of the focussing Nonlinear Schr\"odinger (NLS) equation
which governs the slowly varying envelope of monochromatic waves. 
In this regime, \cite{A} observed that if the phase speed (the speed of
crests of the carrier wave) and the group speed (the speed
of the envelope) are equal, the solitary waves of NLS will describe
approximately solitary waves in the primitive fluid equations. It is simple to show that at
a minimum of the phase-speed, group and phase speed coincide, and
that this occurs at nonzero wavenumber in GC waves for sufficiently
deep water. (The condition is that the Bond number be less than 1/3 which 
roughly corresponds to a depth greater than a centimetre in the air-water problem.) 

In a two dimensional fluid
domain (corresponding to a one dimensional free-surface and
henceforth denoted as the `1D problem'), wavepacket solitary waves
were first computed in the full fluid equations by
\cite{LH} and by \cite{VD}. In three-dimensions (with a two-dimensional free-surface, denoted the
`2D problem'), the first computations of steady solitary waves were
by \cite{PVC1}, with related work by \cite{KA} and \cite{M} on reduced equations. Due to the highly oscillatory and spatially extended
nature of these waves, accurate computations are challenging. Localised waves on a
two-dimensional water surface are often called ``lumps", a
name carried over from their shallow water counterparts
which can be approximated by the localised ``lump" solutions of the KP
equation, but we refer to them as wavepacket solitary waves. \footnote{Naming these waves ``wavepacket" solitary waves is more appropriate since it
differentiates oscillatory solitary waves bifurcating from finite wavenumber from
long waves bifurcating from zero wavenumber.}  In this paper we
focus only on the infinite depth case since for a water-air interface, any GC dynamics in water deeper than a few
centimetres is essentially in the infinite depth regime. In this regime for an air-water interface, the waves bifurcate with a carrier wavelength of approximately  1.7 cm. and a speed of 23 cm./sec.

The stability and dynamics of GC solitary waves in deep water has only recently
been studied. In the 1D problem, \cite{CYA} studied the question of linear stability
and \cite{MVW} studied the time-dependent evolution of solitary waves (stability and
collisions) in full potential flow. In 2D, there are far
fewer studies. The transverse instability of line solitary waves
(solitary waves of the 1D problem trivially extended in the
transverse variable) has been considered by
{\cite{KA2}} and others. Line solitary waves are unstable to transverse perturbations of
sufficiently long wavelength. Fully 2D dynamics have been considered
by  \cite{AM1} in a one-way simplified model and in a quadratic isotropic model in \cite{AM2}. The main goal of this
paper is to study 2D dynamics within a close approximation of the Euler
equations. A simplification that we make, suggested and used in
\cite{KDM} for 1D time-dependent solutions, is to take
a cubic truncation of the scaled Dirichlet-to-Neuman operator that appears
in the free-surface boundary conditions. (One important difference from the model of
\cite{KDM} is that we use the full surface tension term which results in a considerably 
better approximation of the full equations at larger amplitudes.) The approximation proposed is the
simplest model that can \textit{quantitatively} capture small and moderate amplitude
Euler nonlinear CG solitary wave dynamics - a claim we support with 1D
comparisons.

The focussing two-dimensional cubic NLS equation is central to the understanding of the existence and stability of these solitary waves. 
We shall see that this equation can be used to correctly predict the existence and certain instabilities of line and wave packet solitary waves, 
but does not capture the larger amplitude stability characteristics, the asymptotic dynamics of unstable waves nor the interaction of solitary waves. 
(The situation in 1D is worse, since the NLS does not even capture the instabilities of arbitrarily small waves correctly - see for example,
\cite{CYA} and \cite{MVW}.)

This paper is structured as follows: in Section \ref{formulation} we briefly present the derivation of the cubic truncation model we shall use and discuss what can be learned about the solitary waves from the associated NLS equation. In Section \ref{results} we present the numerical results: 1D comparisons between the cubic model and the full problem, followed by 2D bifurcation diagrams, and stability and collision calculations. In Section \ref{extensions} we briefly introduce variations on the cubic model, principally the addition of forcing and dissipation and treating the finite depth problem.

\section{Formulation}\label{formulation}
\subsection{Governing Equations}
Consider the three-dimensional free-surface water wave problem
under the influence of both gravity and surface tension. Let $(x,y)$
denote the horizontal plane, $z$ the vertical direction and $t$
time. The fluid is assumed to be inviscid and irrotational and therefore
there exists a potential function $\phi$, such that the fluid
velocity $(u,v,w)=(\p_x\phi,\p_y\phi,\p_z\phi)$. If the displacement
of the water surface is designated by $z=\eta(x,y,t)$, then the
governing equations for water waves read
\begin{eqnarray}
&&\phi_{xx}+\phi_{yy}+\phi_{zz}=0 \quad\quad\quad\quad\quad\quad
\quad\quad\quad\quad\quad\quad\quad\quad\text{$z<\eta(x,y,t)$} \label{Laplace}\\
&&\big(\phi_x,\phi_y,\phi_z\big)\rightarrow0\quad\quad\quad\quad\quad\quad\quad\quad
\quad\quad\quad\quad\quad\quad\quad\text{as
$|x|+|y|+|z|\rightarrow\infty$} \label{decay}\\
&&\eta_t+\eta_x\phi_x+\eta_y\phi_y-\phi_z=0
\quad\quad\quad\quad\quad\quad\quad\quad\quad\quad\quad\text{at
$z=\eta(x,y,t)$} \label{KBC}\\
&&\phi_t+\frac{1}{2}\big[\phi_x^2+\phi_y^2+\phi_z^2\big]+g\eta=\frac{\sigma}{\rho}
\nabla\cdot\Big[\frac{\nabla\eta}{\sqrt{1+|\nabla\eta|^2}}\Big]
\quad\text{at $z=\eta(x,y,t)$}  \label{DBC}
\end{eqnarray}
where $\nabla=(\p_x,\p_y)$ is the horizontal gradient operator, and
$\nabla\cdot$ is the corresponding horizontal divergence operator.
The constants $g,\rho,\sigma$ are the acceleration due to gravity,
density, and the coefficient of surface tension, respectively.
Following \cite{CS}, who worked in the canonical
variables introduced by \cite{Z1}, the Kinematic and Dynamic boundary conditions (\ref{KBC}-\ref{DBC}) can be recast in terms of
the free surface potential $\xi(x,y,t)=\phi(x,y,\eta(x,y,t),t)$ and $\eta$ as
\begin{eqnarray}
\eta_t&=&G(\eta)\xi\label{zcsa}\\
\xi_t&=&\frac{1}{2(1+|\nabla\eta|^2)}\Big[(G(\eta)\xi)^2-|\nabla\xi|^2+2(G(\eta)\xi)
\nabla\xi\cdot\nabla\eta-|\nabla\xi|^2|\nabla\eta|^2 \nonumber \\
&&+(\nabla\xi\cdot\nabla\eta)^2\Big]-\eta+\nabla\cdot\Big[\frac{\nabla\eta}{\sqrt{1+|\nabla\eta|^2}}\Big] \label{zcsb}
\end{eqnarray}
$G(\eta)$ is a scaled Dirichlet to Neumann (DtN) operator yielding the vertical velocity of the free surface. It is defined by
$G(\eta)\xi=\phi_z-\phi_x\eta_x-\phi_y\eta_y =
\sqrt{1+|\nabla\eta|^2} \; \phi_{n}$, where $\phi_n$ is the
derivative of the potential in the outward normal direction to the
free surface and $\phi$ satisfies (\ref{Laplace}-\ref{decay}) with $\phi(x,y,\eta(x,y,t),t)=\xi$.
These equations have been also been nondimensionalized using a characteristic
lengthscale $L=\Big(\frac{\sigma}{\rho g}\Big)^{1/2}$, a timescale
$T=\Big(\frac{\sigma}{\rho g^3}\Big)^{1/4}$, and a resulting velocity scale
$V=\Big(\frac{\sigma g}{\rho}\Big)^{1/4}$. In cgs units, $g=981
\text{cm/sec}^2$, and $\frac{\sigma}{\rho}$, the ratio of the
surface tension coefficient and density, equals
$73.5\text{cm}^3/\text{sec}^2$ for water.

\subsection{Expansion and Truncation}
\cite{CM} prove that if the $L^\infty$-norm and
Lipschitz-norm of $\eta$ is smaller than a certain constant, then
$G$ is an analytic function of $\eta$. It follows that the DtN
operator can be naturally written in the form of Taylor expansion in
$\eta$, $G=\sum G_i$. For infinite depth, the first three terms of
the Taylor series are given by
{\begin{eqnarray*}
&&G_0=|D|\\
&&G_1(\eta)=D\cdot\; \eta \; D-G_0\; \eta \; G_0\\
&&G_2(\eta)=-\frac{1}{2}\Big[G_0\; \eta^2 \; |D|^2+|D|^2 \; \eta^2\;
G_0-2G_0\;\eta\; G_0\;\eta \; G_0\Big]
\end{eqnarray*}}
where $D=-i\nabla$ and $|D|=(-\Delta)^{1/2}$. Using the cubic truncation of
$G$ and substituted into the kinematic and dynamic boundary conditions closes the evolution problem
\begin{eqnarray}
\eta_t - |D| \xi &=&(G_1+G_2)\xi \label{zcs3a}\\
\xi_t +  (1 - \Delta) \eta &=& \nabla\cdot\Big[\frac{\nabla\eta}{\sqrt{1+|\nabla\eta|^2}} - \nabla \eta \Big]\nonumber\\
&&+\frac{1}{2}\Big[(G_0\xi)\big(G_0\xi-2G_0\eta
G_0\xi-2\eta\Delta\xi\big)-|\nabla\xi|^2\Big].  \label{zcs3b}
\end{eqnarray}
This formulation and approximation has reduced the three dimensional
nonlinear water wave problem to a two dimensional one involving only
the variables on the surface that is computationally, reasonably
simple. In a doubly periodic setting, each term can be efficiently
computed using a pseudospectral method and the fast Fourier
transform (FFT). We shall henceforth call this model the cubic Dirichelet-to-Neumann
Euler equations (cDtNE). Computational methods based on series truncations of the DtN operator have been used
used in a variety of water wave problems and are summarised in detail by \cite{N}. For the 1D problem where the 
full equations can be numerically integrated using a conformal map method we find (see the Results section) that 
the cubic truncation is extremely accurate. The next physically reasonable truncation for this problem, at fifth order, would involve 
substantially more computational resources.

The cDtNE model can also be obtained from the fourth-order truncation of  the kinetic energy part of
the Hamiltonian expression of the surface water wave problem written in terms of the surface potential. The total
energy of the fluid is the sum of kinetic and potential energies
\begin{eqnarray}
H&=&\frac{1}{2}\int
dxdy\int_{-\infty}^\eta\Big(\phi_x^2+\phi_y^2+\phi_z^2\Big)dz+\frac{1}{2}\int\eta^2dxdy\nonumber\\
&&+\int\Big(\sqrt{1+|\nabla\eta|^2}-1\Big)dxdy,
\end{eqnarray}
and an approximate Hamiltonian can be derived by expanding the
energy in powers of the $\eta, \xi$. This takes the form
\begin{eqnarray}
H[\eta,\xi]&\triangleq&\widetilde{H}[\eta,\xi]+O(\eta^3 \xi^2)\nonumber\\
\widetilde{H}[\eta,\xi]&=&\int\frac{1}{2}\xi
(G_0 + G_1 + G_2)\xi+\frac{1}{2}\eta^2+
\Big(\sqrt{1+|\nabla\eta|^2}-1\Big)dxdy \label{zcsH}
\end{eqnarray}
The equations (\ref{zcs3a}) and (\ref{zcs3b}) can be expressed in
canonical form in the sense of \cite{Z1}:
$\eta_t=\frac{\delta\widetilde{H}}{\delta\xi}$ and
$\xi_t=-\frac{\delta\widetilde{H}}{\delta\eta}$. The system has
further physical conserved quantities, of which mass and momentum
\begin{eqnarray}
\text{Mass}&=&\int\eta\; dxdy \label{zcsM}\\
\text{Momentum}&=&\int\xi \nabla \eta \; dxdy, \label{zcsP}
\end{eqnarray}
are used to monitor the global
accuracy of numerical computations with our truncated equations.

\subsection{The Nonlinear Schr\"{o}dinger Equation}
Traditionally, weakly nonlinear wavepackets are
studied using the resulting cubic nonlinear Schr\"{o}dinger
equation (NLS) for the modulational regime of monochromatic waves. It can be derived by
substituting the ansatz:
\begin{eqnarray}
\begin{pmatrix}
\eta \\
\xi
\end{pmatrix}\sim\epsilon
\begin{pmatrix}
A(X,Y,T)\\
B(X,Y,T)
\end{pmatrix}e^{i(kx+ly-\omega t)}+c.c.
+\epsilon^2\begin{pmatrix}
\eta_1\\
\xi_1
\end{pmatrix}
+\epsilon^3\begin{pmatrix} \eta_2\\
\xi_2
\end{pmatrix}+\cdot\cdot\cdot
\end{eqnarray}
into (\ref{zcsa}), (\ref{zcsb}) and ensuring that the series is
well-ordered for $t=O(\epsilon^{-2})$. Here
$X=\epsilon\big(x-c_gt\big)$, $Y=\epsilon y$ and $T=\epsilon^2t$
where $c_g$ is the group velocity in the wave propagating direction
and ``c.c." represents the complex conjugate of preceding terms. The
wave envelope $A$ can then be found to satisfy the NLS equation (\cite{AS})
\begin{eqnarray}
iA_T+\lambda_1A_{XX}+\lambda_2A_{YY}=\mu|A|^2A \label{NLS}
\end{eqnarray}
It is a trivial fact that substituting the same ansatz into the
equations  (\ref{zcs3a}), (\ref{zcs3b}) instead yields an NLS
equation with identical coefficients - which is the motivation
for choosing at least a cubic truncation of the DtN operator. We omit the
details of the derivation, and just state the results. Choosing
$k=1,l=0$ as the carrier wave, the phase and group velocity are
equal, with the phase velocity at its minimum $c_{min}=\sqrt{2}$. All of the waves we consider bifurcate from this point and exist only for $c<c_{min}$.
The coefficients of NLS are given by
\[
\lambda_1 = \frac{\sqrt{2}}{4}, \qquad \lambda_2 = \frac{\sqrt{2}}{2}, \qquad \mu = -\frac{11}{8}\sqrt{2}.
\]
The solution to the original system is then as follows
\begin{eqnarray}
&&\eta=\epsilon
Ae^{i\Theta}-2\epsilon^2A^2e^{2i\Theta}+\cdot\cdot\cdot+c.c.\\
&&\xi=-i\sqrt{2}\epsilon
Ae^{i\Theta}+i2\sqrt{2}\epsilon^2A^2e^{2i\Theta}+\cdot\cdot\cdot+c.c.
\end{eqnarray}
where $\Theta=x-\sqrt{2}t$. Since the NLS equation (\ref{NLS}) is of
the elliptic or focussing type with solitary wave solutions in both
one and two dimensions, and since the phase and group speed are
equal at the chosen carrier wave, one can expect small amplitude
solitary waves bifurcating from a uniform flow. These solitary waves
of (\ref{zcs3a},\ref{zcs3b}) can be approximated by the NLS solitary
waves found by solving the elliptic eigenvalue problem for
$\rho(x,y)$ and $\Omega$
\begin{eqnarray}
\Delta\rho+\rho^3=\Omega\rho, \qquad ||\rho||_\infty=1, \qquad \rho \rightarrow 0 \text{ at } \infty.  \label{nonlinEV}
\end{eqnarray}
This is obtained by setting
$A=|\mu|^{-1/2}\rho(|\lambda_1|^{-1/2}X,|\lambda_2|^{-1/2}Y)e^{-i\Omega
T}$. Denoting $\Delta c\equiv\sqrt{2}-c$ where c is the wave
propagating speed, then, by the chosen scaling, $\Delta
c\sim\Omega\epsilon^2$. The total energy of the Gravity-Capillary
solitary wave bifurcating below the minimum phase speed is then calculated
by $(2.10)$. 

For the one dimensional problem (\ref{nonlinEV}) has the well known unique focussing NLS soliton, $\rho=\text{sech}(x)$,
$\Omega=\frac{1}{2}$, $\int\rho^2=2\sqrt{2}$ and thus
\begin{eqnarray}
&&||\eta||_\infty\thickapprox\Big|\frac{8}{\mu}\Big|^{1/2}\big(\Delta
c\big)^{1/2}+\Big|\frac{8}{\mu}\Big|\Delta c\\
&&\widetilde{H}\thickapprox\frac{4|\lambda_1|^{1/2}}{|\mu|}\int\rho^2\Big(\frac{\Delta
c}{\Omega}\Big)^{1/2}=\frac{2^{23/4}}{11}\big(\Delta c\big)^{1/2} \label{1Dbif}
\end{eqnarray}

\begin{figure}
\begin{center}
\psfig{file=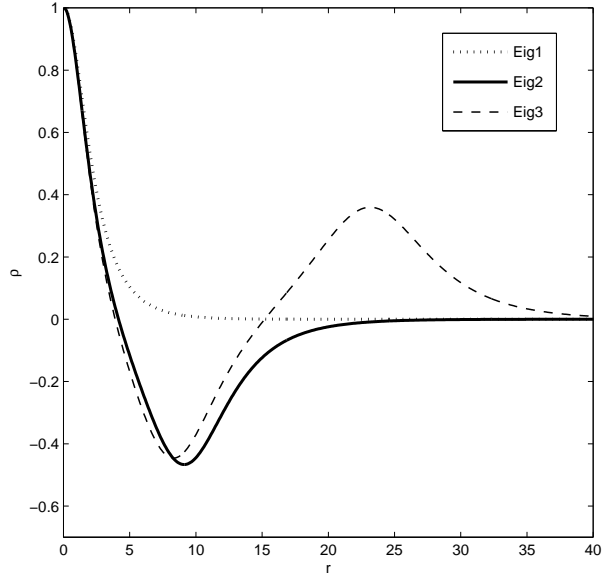,width=0.6\textwidth} 
\end{center}
\caption{First three radially symmetric solutions of (\ref{nonlinEV}). The corresponding energies of the wavepacket solitary waves for these three envelopes are 12.04, 79.39, 201.46} \label{FigNonlinEV}
\end{figure}

For the two dimensional case, which are the ones of interest here, there are countably many solutions to
the problem (\ref{nonlinEV}) (see \cite{AEKLS} and references therein), and the first three radially
symmetric solutions are shown in Figure \ref{FigNonlinEV}. For the
ground state, called the Townes soliton in nonlinear optics (\cite{CGT}), $\Omega\thickapprox0.204$,
$\int\rho^2\thickapprox11.70$, and, in the present hydrodynamic context,
\begin{eqnarray}
&&||\eta||_\infty\thickapprox\Big|\frac{4}{\mu\Omega}\Big|^{1/2}\big(\Delta
c\big)^{1/2}+\Big|\frac{4}{\mu\Omega}\Big|\Delta c\\
&&\widetilde{H}\thickapprox\frac{4|\lambda_1\lambda_2|^{1/2}}{|\mu|}\int\rho^2\thickapprox12.04
\end{eqnarray}

The details of these computations can be found in \cite{AM1}. An
interesting particularity is that the physical energy for two dimensional
wavepacket solitary waves is predicted to tend to a \textit{finite} value
of 12.04 as the amplitude approaches zero, which we shall verify in
the cDtNE model. This is purely a consequence of scaling: the
radially symmetric envelope's area is proportional to
$\epsilon^{-2}$ exactly countering the effect of decreasing
amplitude of the wave whose energy density is $\epsilon^{2}$.
Shallow water lump solutions of KP (a valid approximation when the Bond number is greater than 1/3) do not have this property. The higher energy states of the eigenvalue problem (\ref{nonlinEV}) can be
associated with different families of travelling waves of the cDtNE
model resulting in a remarkable ``quantisation" of the energy of
solitary GC waves in the 2D problem.

\section{Results}\label{results}
The numerical solution of the cDtNE system is implemented on a
periodic domain with Fourier pseudo-spectral methods, where all
derivatives and Hilbert transforms are computed in Fourier space
with spectral accuracy, while nonlinearities are computed
pseudo-spectrally in real space. For traveling waves, the resulting
algebraic system for the Fourier coefficients is solved using
Newton's method using a monochromatic wave modulated by the solution
to (\ref{nonlinEV}) as initial guesses. The branches are computed
through straightforward continuation methods. For time integration
of the system, a classic fourth order Runge-Kutta method is used
with the integrating factor method (see \cite{MT} for details) used
to exactly integrate the linear part of the equation. The conserved
quantities of the system are monitored and in all cases are
preserved to a relative error of at most $O(10^{-4})$. All the
computations are de-aliased with a doubling of Fourier modes. For
two dimensional computations at least $256\times64$ modes are used
along the propagating and transverse directions respectively. It is
often the small amplitude solutions that are most difficult to
compute accurately since the spatial decay of those solutions is
much slower. Thus, the computational the domain is gradually
enlarged as the amplitude becomes smaller.

\subsection{The 1D problem and model accuracy}

\begin{figure}
\begin{center}
\psfig{file=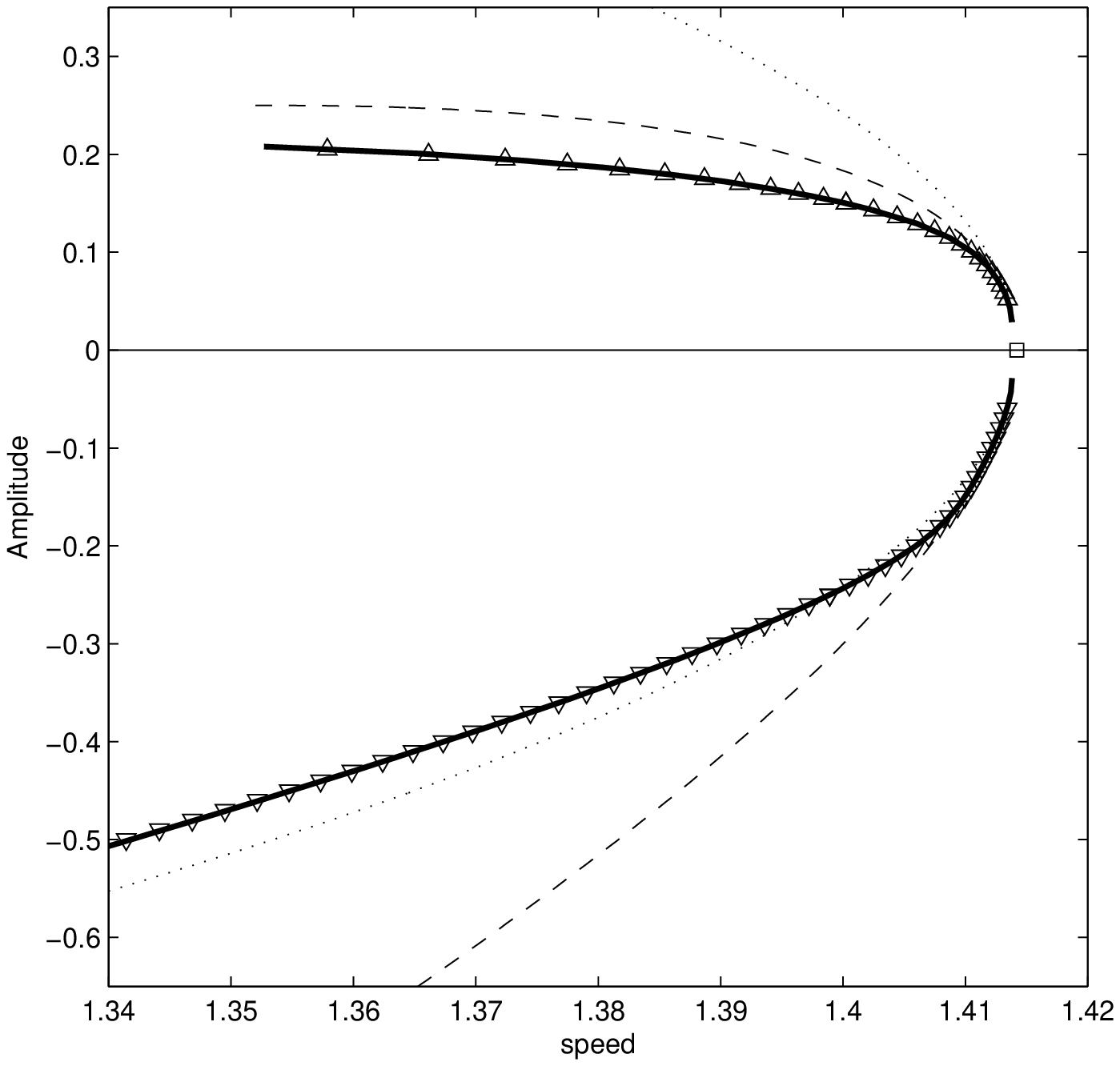,width=0.48\textwidth} \quad
\psfig{file=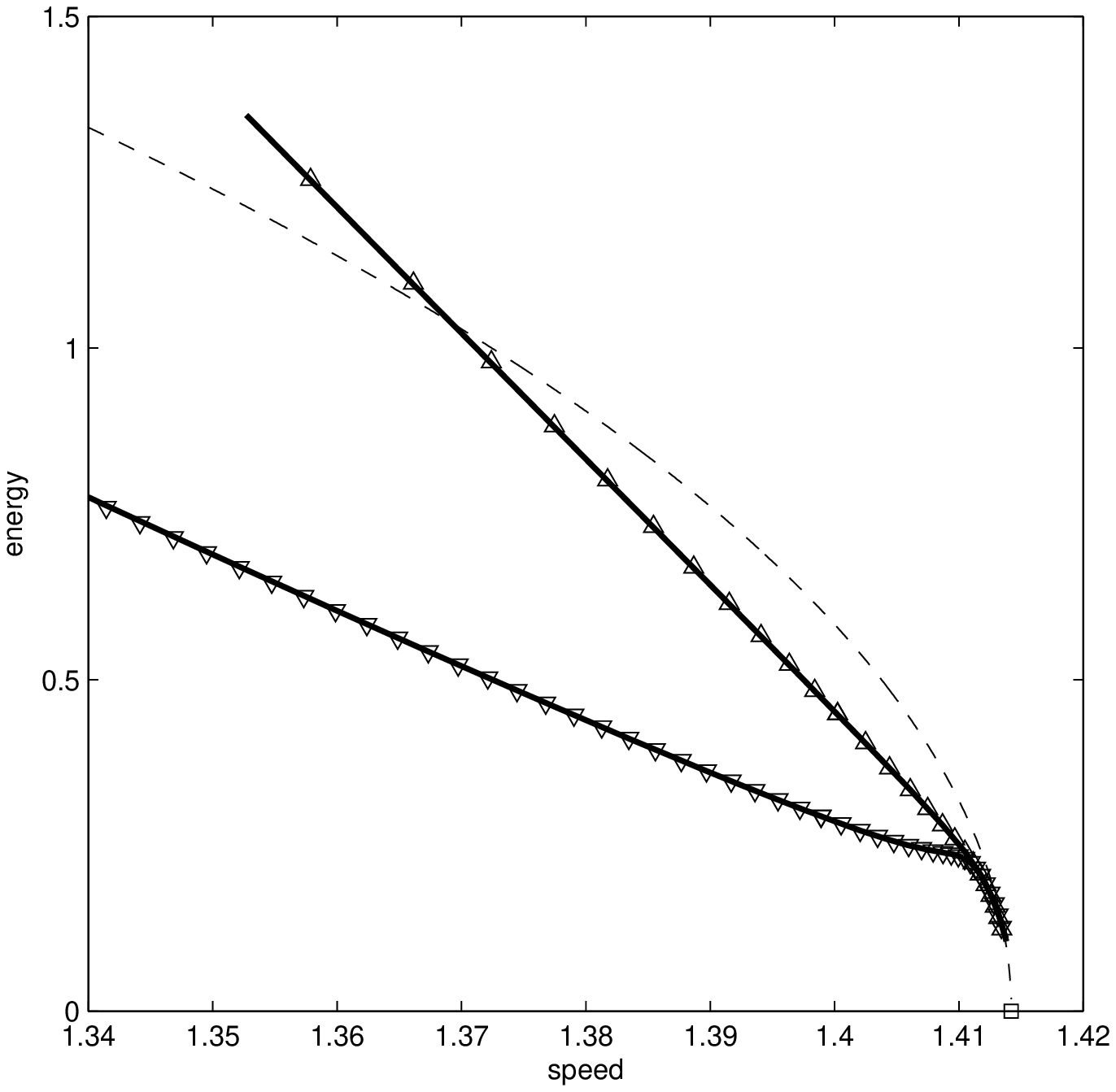,width=0.48\textwidth}
\end{center}
\caption{Left: 1D speed-amplitude bifurcation
picture for elevation and depression solitary waves.
The solid line, triangles, dotted lines and  dashed lines correspond to cDtNE, Euler, leading order NLS and corrected NLS (from equation \ref{1Dbif})
respectively. Right: 1D speed-energy bifurcation
picture for elevation and depression
solitary waves. The triangles denote the Euler solutions, the solid line is the cDtNE, and the dashed
line is the prediction of NLS.} \label{fig1}
\end{figure}

\begin{figure}
\begin{center}
\psfig{file=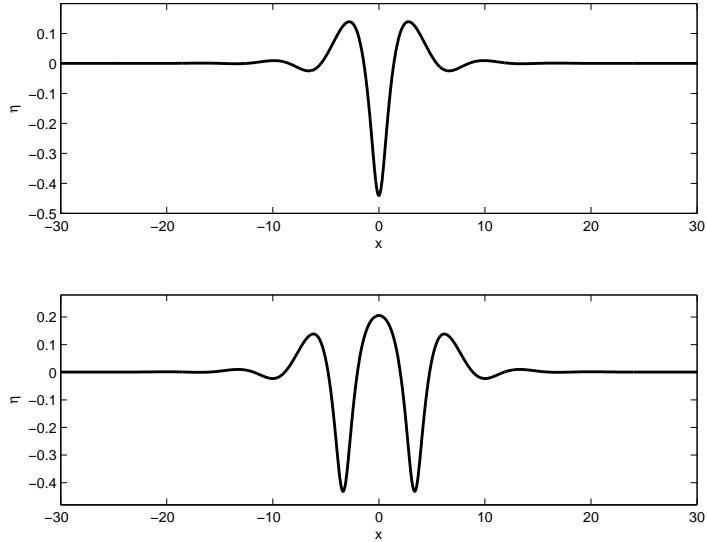,width=0.7\textwidth}
\end{center}
\caption{Typical free-surface solitary profiles of 1D elevation and depression solitary waves for the cDtNE model far from the NLS wavepacket regime. Top: depression wave with $\eta(0)=-0.4406$ and $c=1.3573$. Bottom: elevation wave with $\eta(0)=0.2055$ and $c=1.3579$. The solution to the full equations {\sl cannot} be visually differentiated from these on this scale. For the depression wave, the maximum difference between the cDtNE and full solutions is $6 \times 10^{-4}$.} \label{fig2}
\end{figure}

\begin{figure}
\begin{center}
\psfig{file=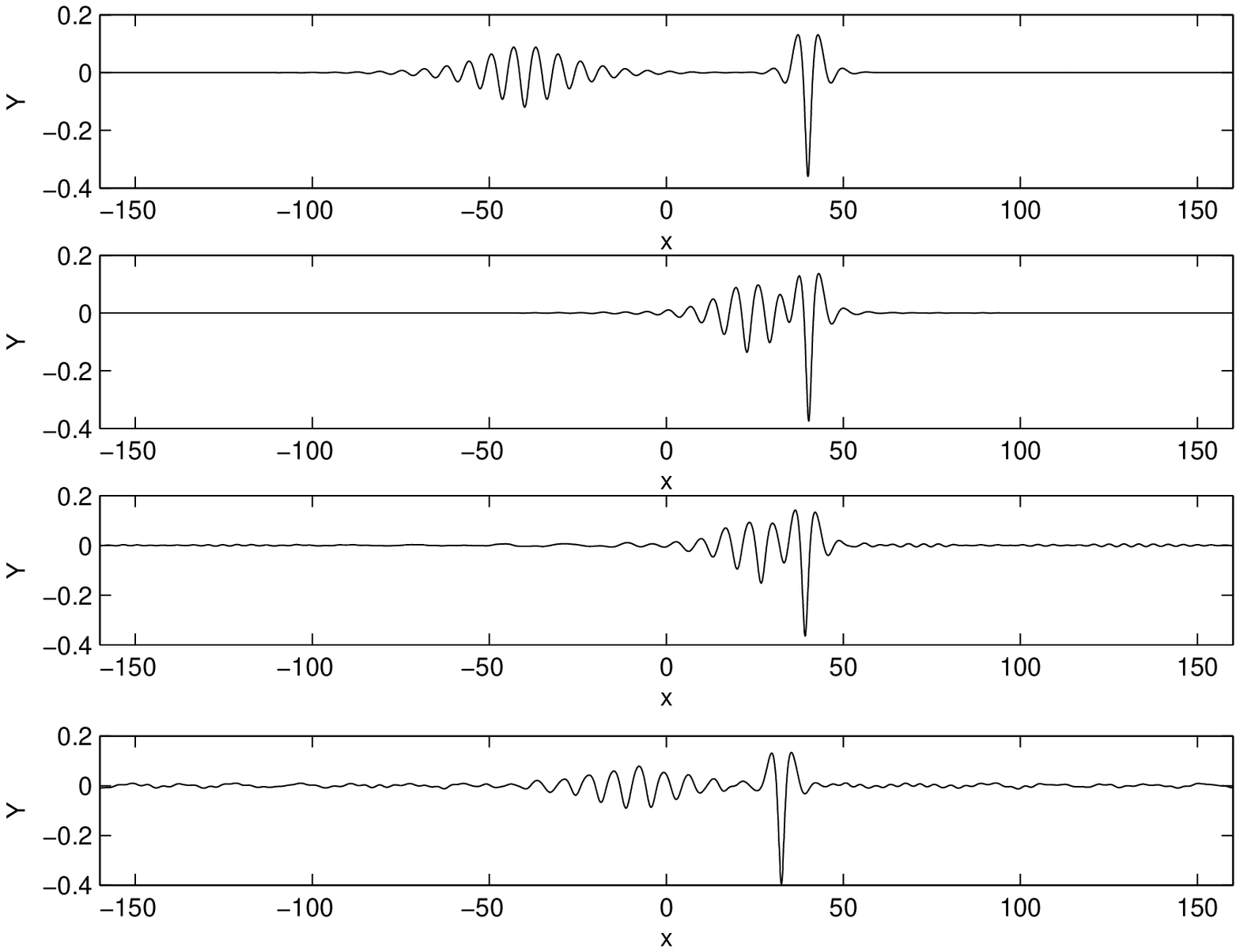,width=0.46\textwidth} \quad
\psfig{file=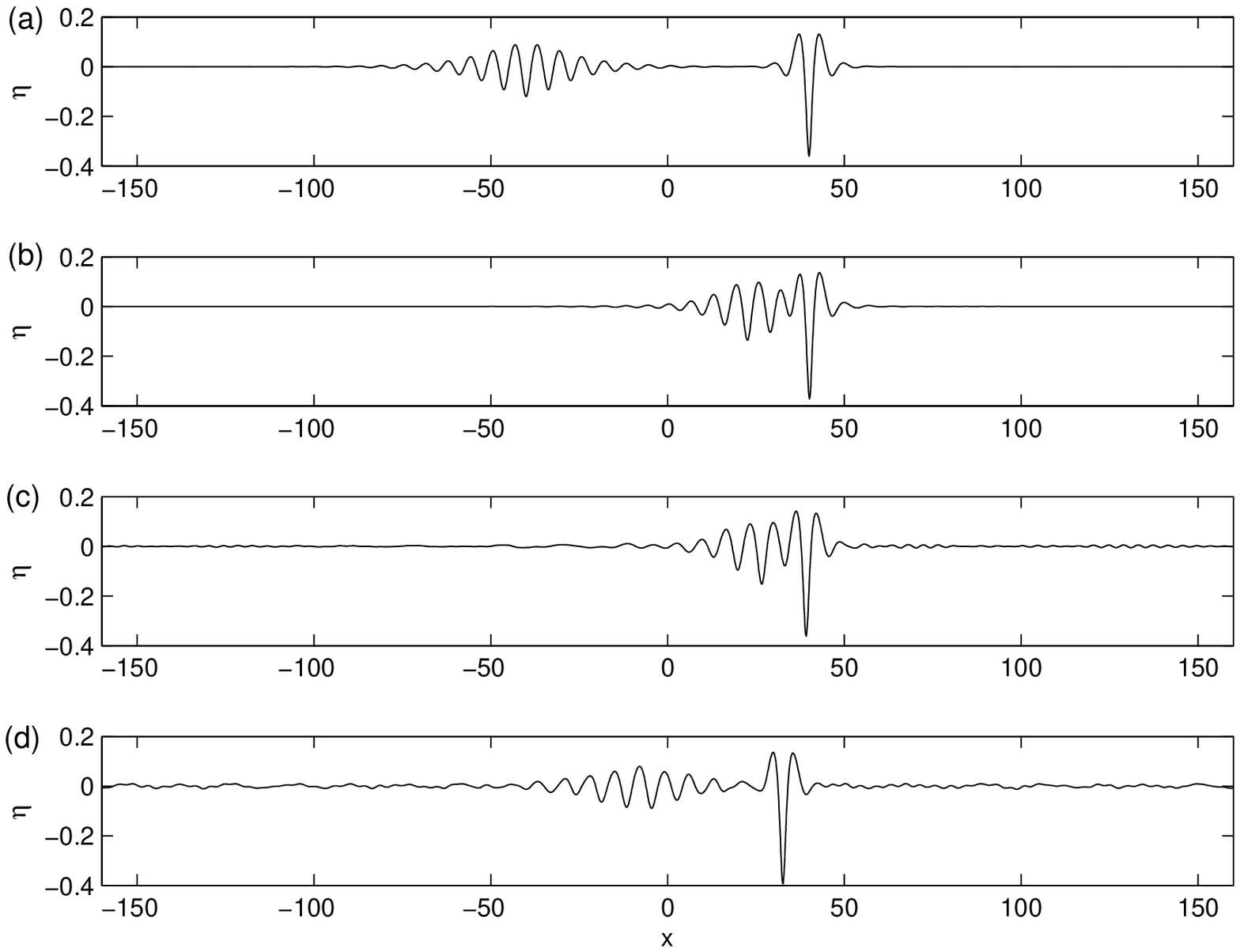,width=0.49\textwidth}
\end{center}
\caption{Overtaking collision of two depression waves of different amplitudes (minimum free surface heights of -0.36 and -0.12) shown in a frame of reference moving at the speed of the larger wave. From top to bottom: $t=0$, $t=2000$, $t=2500$ and $t=3500$. Only the larger wave survives the collision, with the smaller wavepacket, to the left of it eventually dispersing. Left: Full equations from \cite{MVW}. Left: cDtNE} \label{figure1DTDcomp}
\end{figure}

The bifurcation diagrams of 1D
Gravity-Capillary solitary waves for the full equations in deep
water have been presented in \cite{VD} and others. In figure \ref{fig1} we compare
the speed-amplitude and speed-energy bifurcation diagrams of the cDtNE model to these and to the
theoretical prediction of NLS. There, and elsewhere, we use either $\eta$ at
the centre of the wave  or the energy $\widetilde{H}$ as the amplitude parameter. The cDtNE model
agrees well with the bifurcation picture for the Euler equations far
beyond the NLS regime. Typical profiles of the depression and
elevation branches of waves are show in figure \ref{fig2}. The model
is remarkably accurate: at relatively large amplitudes which are far
from the NLS regime the relative difference in profile between the
full equations and the model are of order $10^{-3}$ and not visually
discernible. Although a quantitative comparison of time-dependent dynamics is involved and beyond the
scope of this paper (in particular methods used to solve the full equations usually do not use a uniform grid in $x$), we show an example of the inelastic overtaking collision of two solitary waves computed with the full equations (reported in \cite{MVW}) and cDtNE truncation in Figure \ref{figure1DTDcomp}. The results are extremely close and we see this as further evidence that the cDtNE model is an accurate representation of the Euler equations in a broad amplitude range of the CG regime. Comparisons that we have made with the cubic model of \cite{KDM} show that retaining the full nonlinearity of the surface tension term is far more important than the nonlinearity associated with higher order corrections of the Dirichelet to Neumann map. One may conjecture that the surface tension term has a strong regularising effect that implies fast convergence of the DtN power series approximation throughout the evolution.

\subsection{The 2D bifurcation problem}

\begin{figure}
\begin{center}
\psfig{file=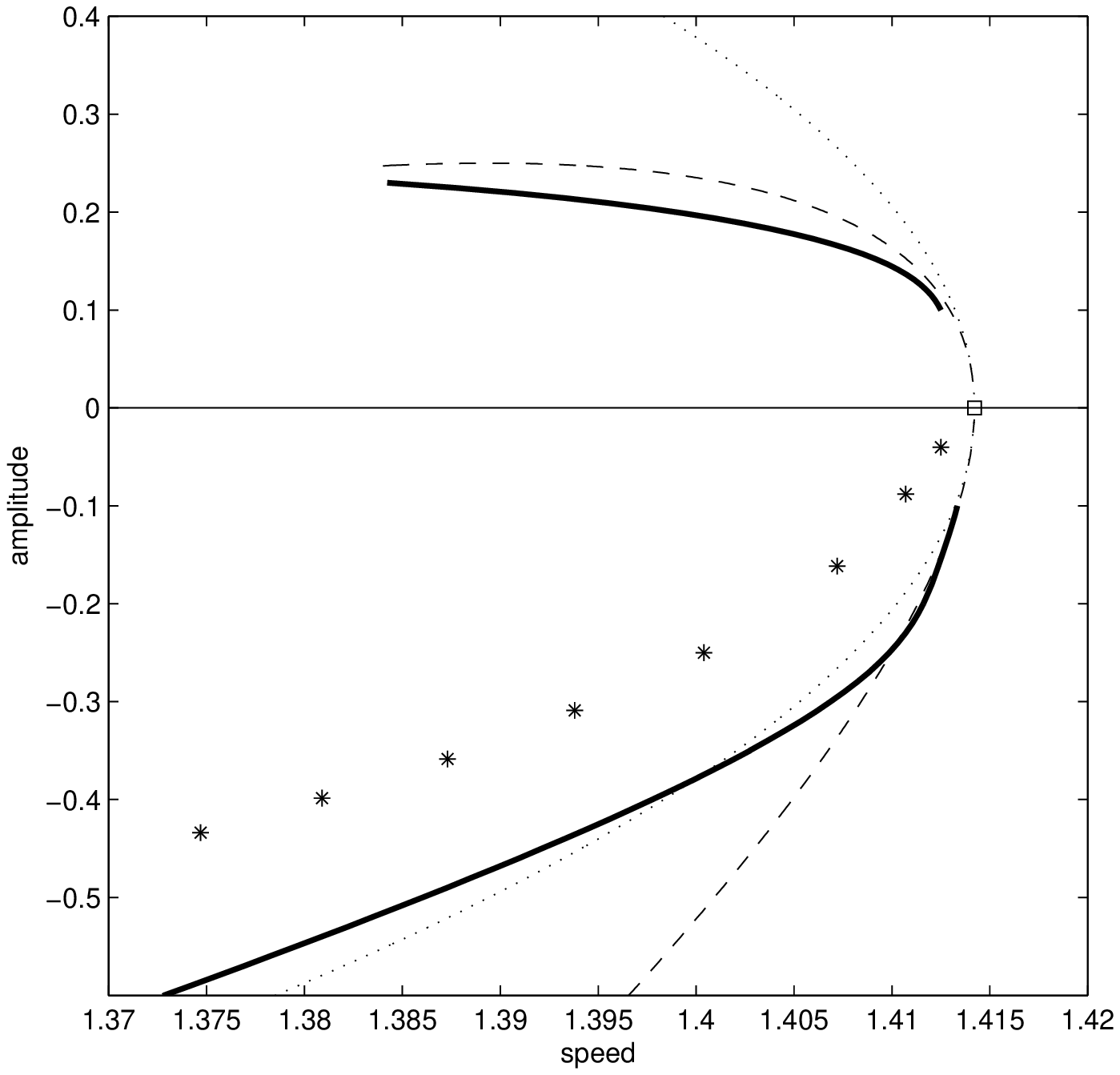,width=0.48\textwidth} \quad
\psfig{file=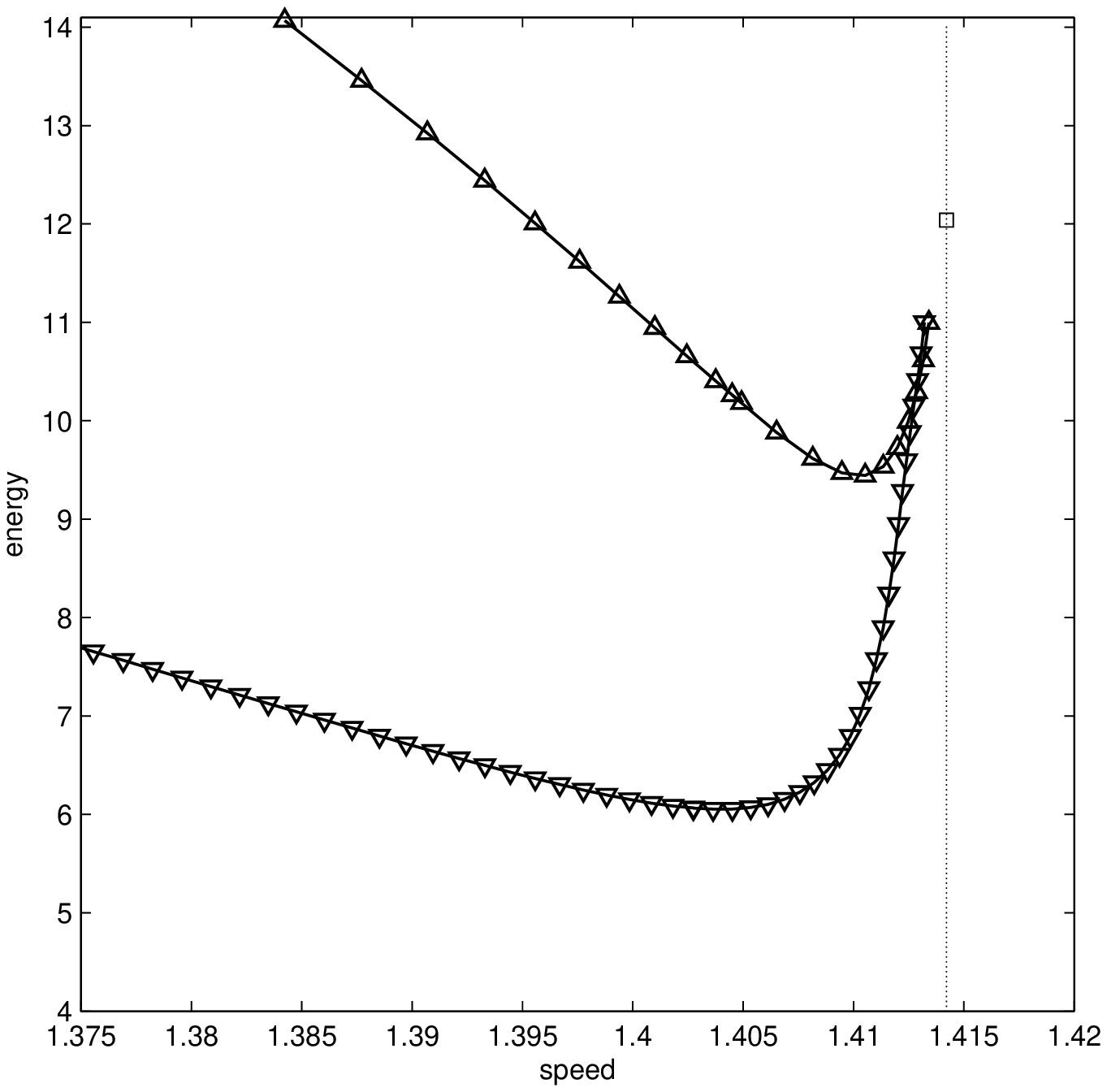,width=0.48\textwidth}
\end{center}
\caption{Left: Speed-amplitude bifurcation curves for elevation and depression solitary waves. The cDtNE, Euler, NLS and NLS plus correction are shown respectively by the solid, stars, dot and dash lines. Right: speed-energy bifurcation picture for both elevation (triangles pointing up) and depression (triangles pointing down). The bifurcation energy predicted by NLS is indicated by a square.}
\label{figBif2D}
\end{figure}

\begin{figure}
\begin{center}
\psfig{file=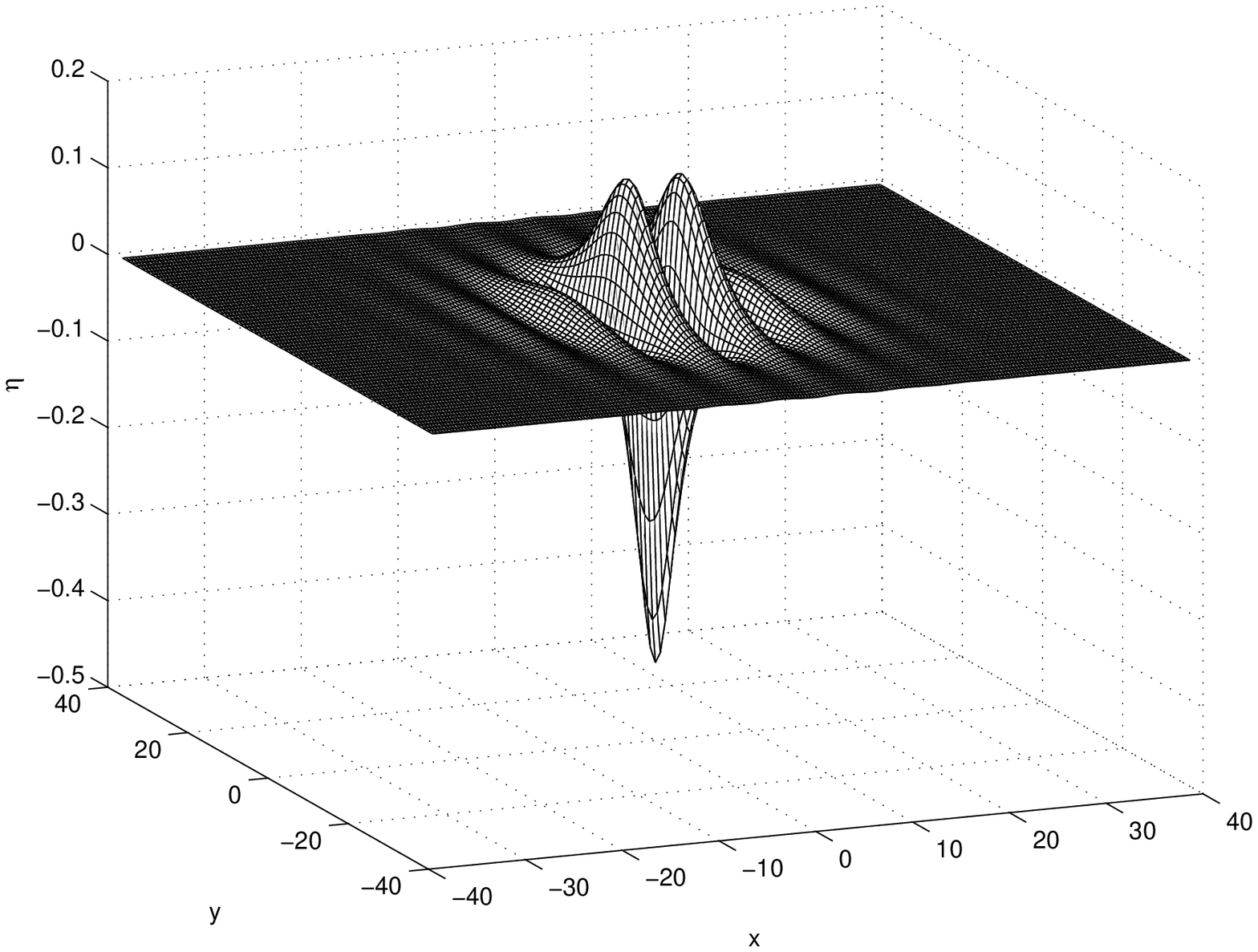,width=0.48\textwidth} \quad
\psfig{file=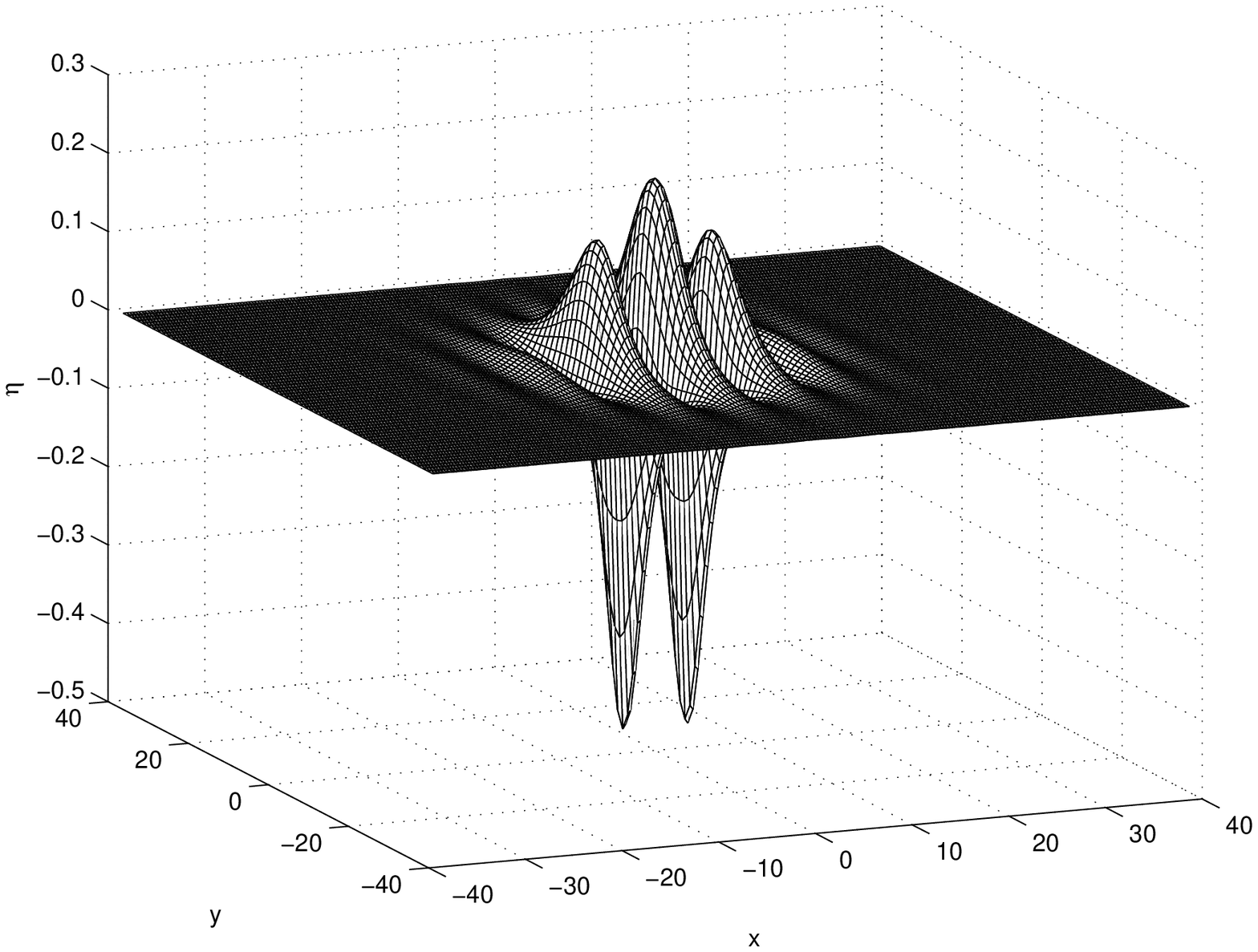,width=0.48\textwidth}
\end{center}
\caption{Top: representative of the depression wavepacket solitary
wave solution to the cDtNE equation with $\eta(0,0)=-0.41$ and
$c=1.3907$. Bottom: representative of the elevation wavepacket solitary
wave solution to the cDtNE equation with $\eta(0,0)=0.23$ and
$c=1.3842$.} \label{figProfiles2D}
\end{figure}

Two-dimensional GC solitary waves of the full equations in infinite
depth were first computed in
{\cite{PVC1}} using finite
differences and boundary integral methods. A more resolved computation
was performed by P$\breve{a}$r$\breve{a}$u and reported in
\cite{CDAD2}. However, this later computation is, in our opinion,
also under-resolved, particularly at small amplitudes where the waves are highly oscillatory and spatially extended, as it deviates
considerably from both the cDtNE and the NLS results.  Figure
\ref{figBif2D} shows the bifurcation diagram of GC solitary waves,
as obtained from numerical solutions of the cDtNE, full potential
flow (as computed by P$\breve{a}$r$\breve{a}$u and reported in \cite{CDAD2}), and both the leading order
and leading order plus first correction of the NLS approximation. We
conjecture that the cDtNE would be in quantitative agreement with
full potential flow in the 2D problem given the accuracy of results
at moderate amplitude in the 1D case, and the agreement with NLS
approximations at small amplitude. The discrepancy
between the full Euler results and the expected amplitudes predicted
by NLS had been noted in \cite{CDAD2}. Furthermore, we note that the
cDtNE wave packets have finite energy at their bifurcation point, as
predicted by NLS but which cannot be verified in the full Euler
computations at current resolution (P$\breve{a}$r$\breve{a}$u, private communication). Typical
profiles of the elevation and depression waves corresponding to the
bifurcation curves presented in Figure  \ref{figBif2D}  are shown in
Figure \ref{figProfiles2D}.

All solutions presented so far are those whose profile envelope at
bifurcation is described by the ground state eigenfunction of
(\ref{nonlinEV}). In the 2D problem, we have also computed cDtNE solitary
wave solutions whose envelope is approximated by radially symmetric higher modes. For
example in Figure \ref{FigMode2} we show the unstable evolution of
the complicated travelling wave resulting from the solitary wave
assigned to the second radial mode of (\ref{nonlinEV}) (the wave is
shown in the top-left panel of Figure \ref{FigMode2}). This wave has an
energy of 79.39 at bifurcation, much higher than the ground state. Non-radially symmetric solutions to (\ref{nonlinEV}) are discussed in \cite{AEKLS}, but we did not attempt to compute solitary waves whose envelopes are not radially symmetric.

\subsection{Stability, focussing and wave collapse}

There are a few known results that guide us in a numerical study of
2D stability problem. First, it is known that both depression and
elevation plane solitary waves (1D waves extended in the second
dimension) are linearly unstable with respect to sufficiently long
small perturbations in the transverse direction. This has been shown
both within an NLS approximation
{\cite{RR}} and using arguments based
on linearization of the full equations
{\cite{KA2}}. Second, in 2D the
underlying focussing NLS equation (\ref{NLS}) is well known to
exhibit a finite-time focussing blowup called wave collapse when, in
an unbounded setting, the initial conserved energy $E= \int|\nabla
A|^2 - \frac12 |A|^4 $, is negative
(see {\cite{Z2}} and \cite{SS}). (Note that this energy
is not the same one arising from the cDtNE equations.) The result is
obtained by a virial argument, whence $M_0= \int |A|^2$ is conserved
and $M_2(t) = \int (X^2 + Y^2) |A|^2$ satisfies $M'' = 8E$. Solitary
waves which are solutions to the eigenvalue problem (\ref{nonlinEV})
are critical with $E=0$. Thus, a small negative energy perturbation
leads to focussing and blowup and a positive energy perturbation leads to
spreading of the underlying wave envelope. We thus expect, and will confirm numerically, the
instability of localised solitary wave solutions for nonzero
envelope energy perturbations in the near-NLS limit of the cDtNE
model (where solitary wave envelopes are well approximated by the
NLS).  We note that the simple virial argument is not available in a
periodic setting (see {\cite{SS}}) but
nevertheless seems to predict stability very well in our
computations on periodic domains. Third, both elevation and
depression wave branches have critical point in their speed-energy
relation (see Figure \ref{figBif2D}), which can lead to an exchange
of linear stability of the eigenfunction of
the linearized problem associated to the translational invariance
symmetry  (see, for example,
{\cite{AC}}). This is the instability
commonly observed in small amplitude elevation waves of the 1D
problem leading to the eventual development of a depression wave
\cite{MVW}. 

All figures of solitary wave dynamics presented are
shown is a frame moving with the speed of the wave that was used
construct the initial data. In this frame the main features of the
wave evolve slowly.

\begin{figure}
\begin{center}
\psfig{file=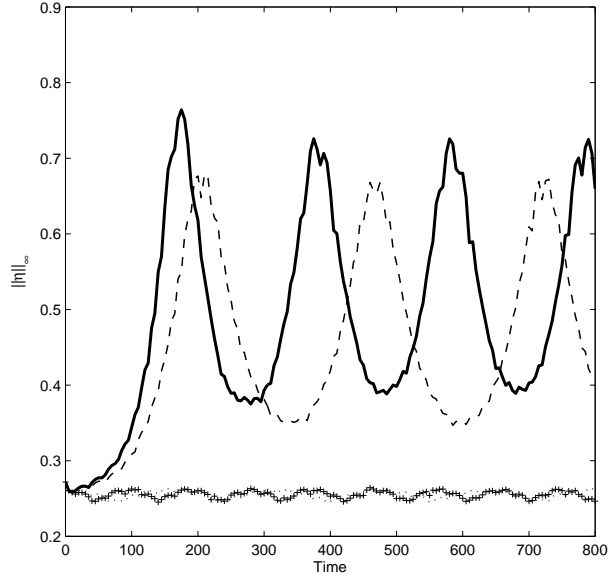,width=0.6\textwidth}
\end{center}
\caption{Transversal instability of the plane solitary wave, $c=1.3994, \eta(0)=-0.247$. The transverse perturbation is obtained by taking an initial data of the form $\eta(x,y,0)=[1 + 0.1 \cos(a l_c y)]\widetilde{\eta}(x)$, where $\widetilde{\eta}(x)$ is a travelling wave solution in 1D, $l_c=18^{1/4} \Delta c^{1/2}$, and $a=0.8$ (solid line), $a=0.9$ (dashed line), $a=1.1$ (dotted line), $a=1.2$ (plus line)} \label{FigTransIns}
\end{figure}

\begin{figure}
\begin{center}
\psfig{file=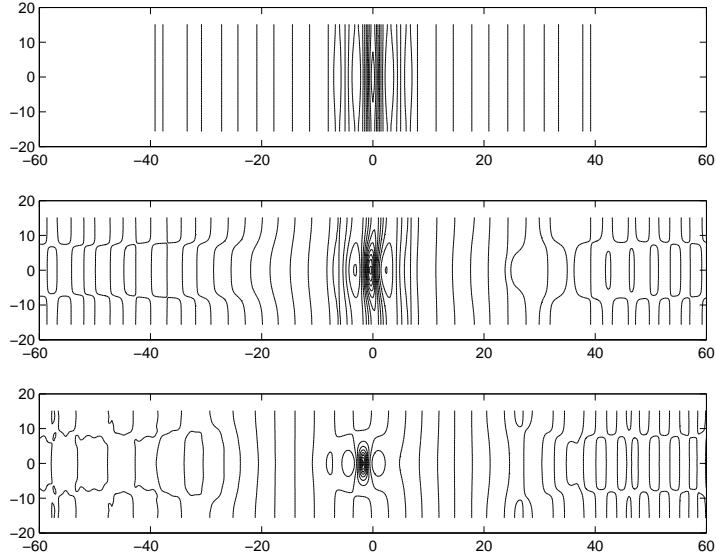,width=0.7\textwidth}
\end{center}
\caption{Evolution of the transverse instability to a localised solution in the case of $a=0.8$ in Figure \ref{FigTransIns}. The contours are shown from top to bottom at $t=0$, $t=125$, $t=175$.} \label{FigTransEvol}
\end{figure}

For plane solitary waves, the linear analysis based on NLS (see
\cite{AM1}) shows that the transverse perturbation
$e^{ily}$ is unstable, when the wave number $l$ in the $y$ direction
satisfies
\begin{eqnarray}
|l|\lesssim\Big|\frac{3\mu}{8\lambda_2}\Big|^{1/2}||u||_{\infty}\thickapprox(18)^{1/4}(\Delta
c)^{1/2}\triangleq l_c
\end{eqnarray}
We confirm this in our numerical experiments where a plane solitary
wave is perturbed with four transverse perturbations of different
wavelengths and the subsequent nonlinear evolution is compared. The
perturbation wavenumbers  are $l=1.2l_c,1.1l_c, 0.9 l_c$, and
$0.8l_c$ respectively. As shown in Figure \ref{FigTransIns} the
first two perturbations do not destabilise the plane wave whereas
the next two do.  The subsequent evolution of the instability shows
a focussing behaviour  reminiscent of the underlying collapse
dynamics of NLS  as shown in Figure \ref{FigTransEvol}. This
intermediate time focussing is arrested by the generation of a
travelling \textit{breather}: a propagating periodic-in-time localised
structure. This structure is best described as a localised
depression solitary wave with periodic amplitude modulation, and
which appears to be stable (see evolution for $t>175$ in Figure
\ref{FigTransIns}) in all our experiments. These breathers are very
common and also occur in the nonlinear evolution of  instabilities
of small amplitude fully localised solitary waves below. 

Throughout our computations in this section,
instabilities and wave interactions will invariably lead
to some high frequency dispersive radiation and thus, due to our use of a
periodic domain, the remaining coherent structures are embedded in a
``sea" of linear ripples. This is visible in most computations and
can also be seen as further evidence of the stability of the
resulting structures.

\begin{figure}
\begin{center}
\psfig{file=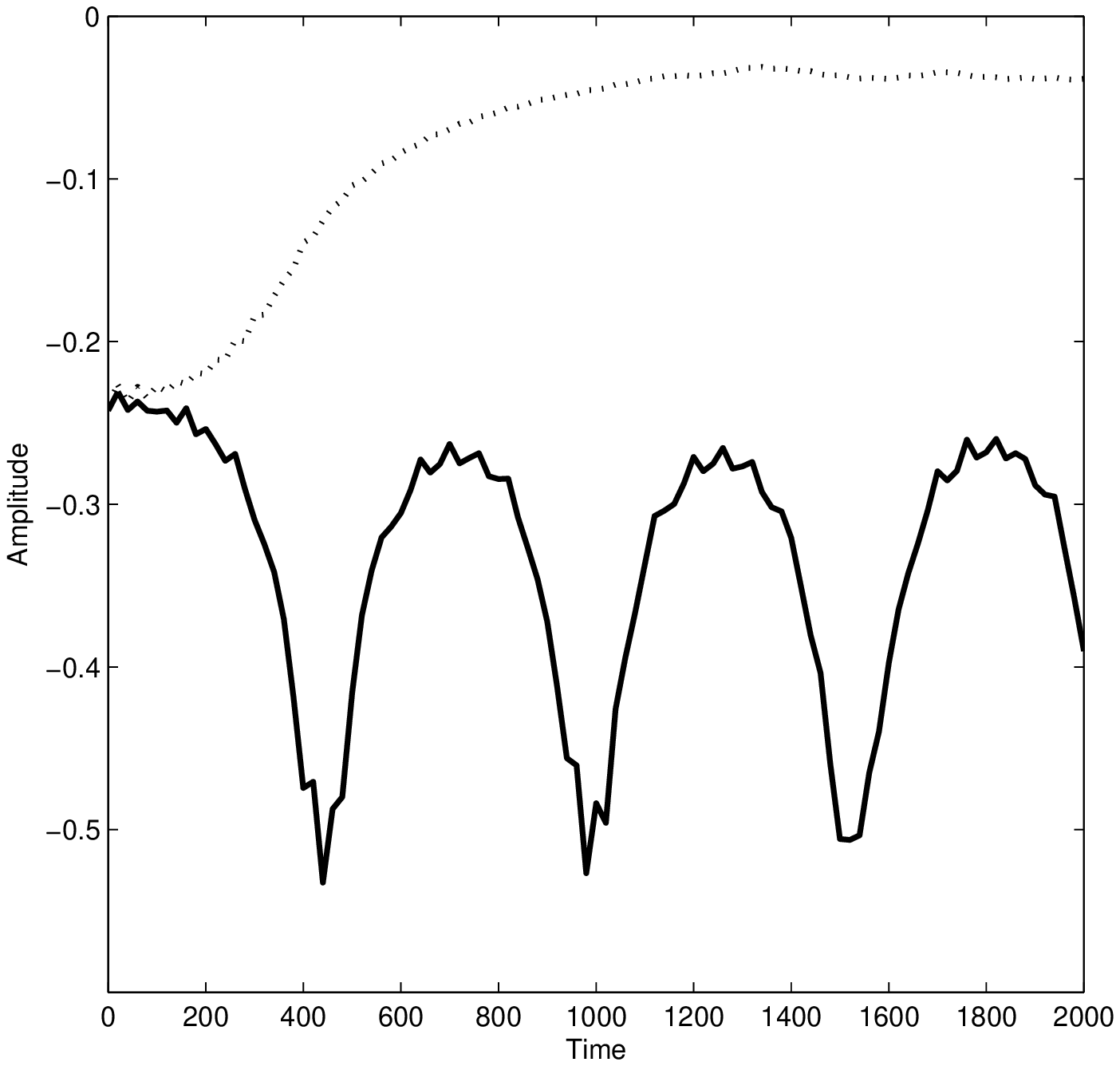,width=0.48\textwidth} \quad
\psfig{file=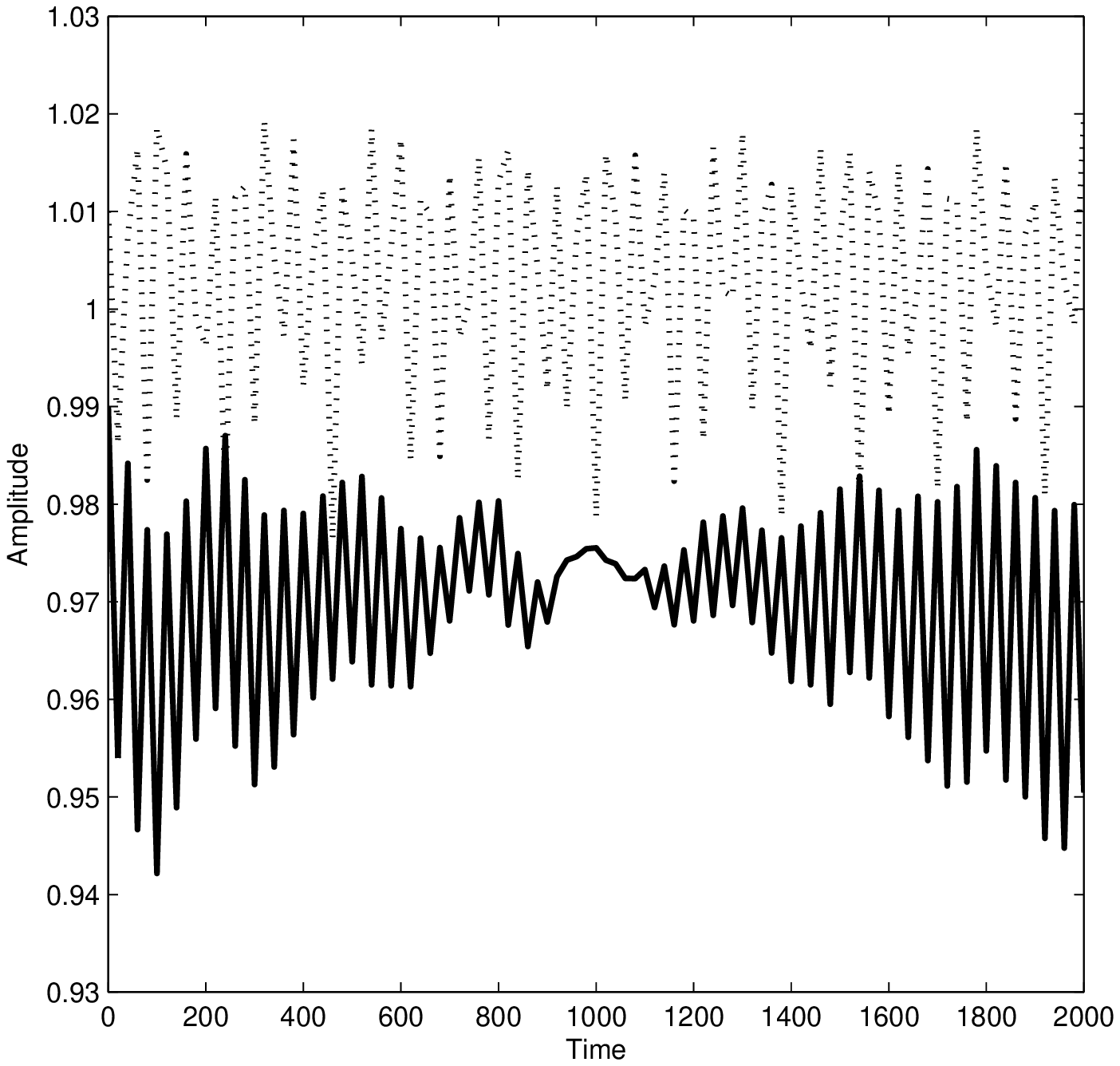,width=0.48\textwidth}
\end{center}
\caption{Left: evolution of maximum trough depth for perturbed
depression waves in the small amplitude regime ($\eta(0,0)=-0.24$ and $c=1.4103$). Negative energy perturbation ($\delta=0.01$) is shown by a solid
line, and positive energy  ($\delta=-0.01$) by a dashed line.
Right: evolution of maximum trough depth (normalised by initial maximum trough depth) for
depression waves in the large amplitude regime ($\eta(0,0)=-0.49$ and $c=1.3873$). Positive energy perturbation ($\delta=0.01$) is shown by a solid
line, and negative energy  ($\delta=-0.01$) by a dashed line.}
\label{DepLumpStab}
\end{figure}

\begin{figure}
\begin{center}
\psfig{file=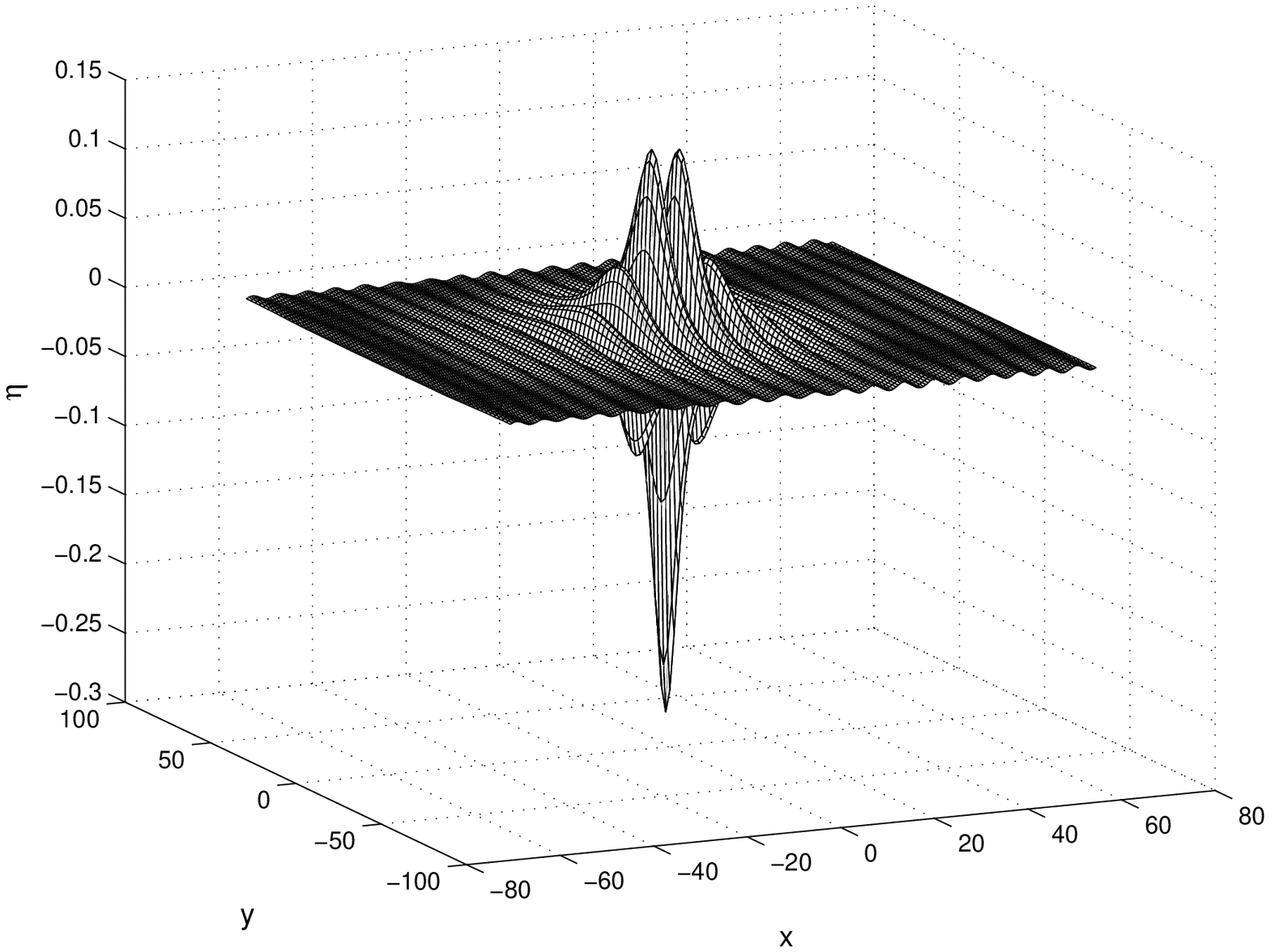,width=0.48\textwidth} \quad
\psfig{file=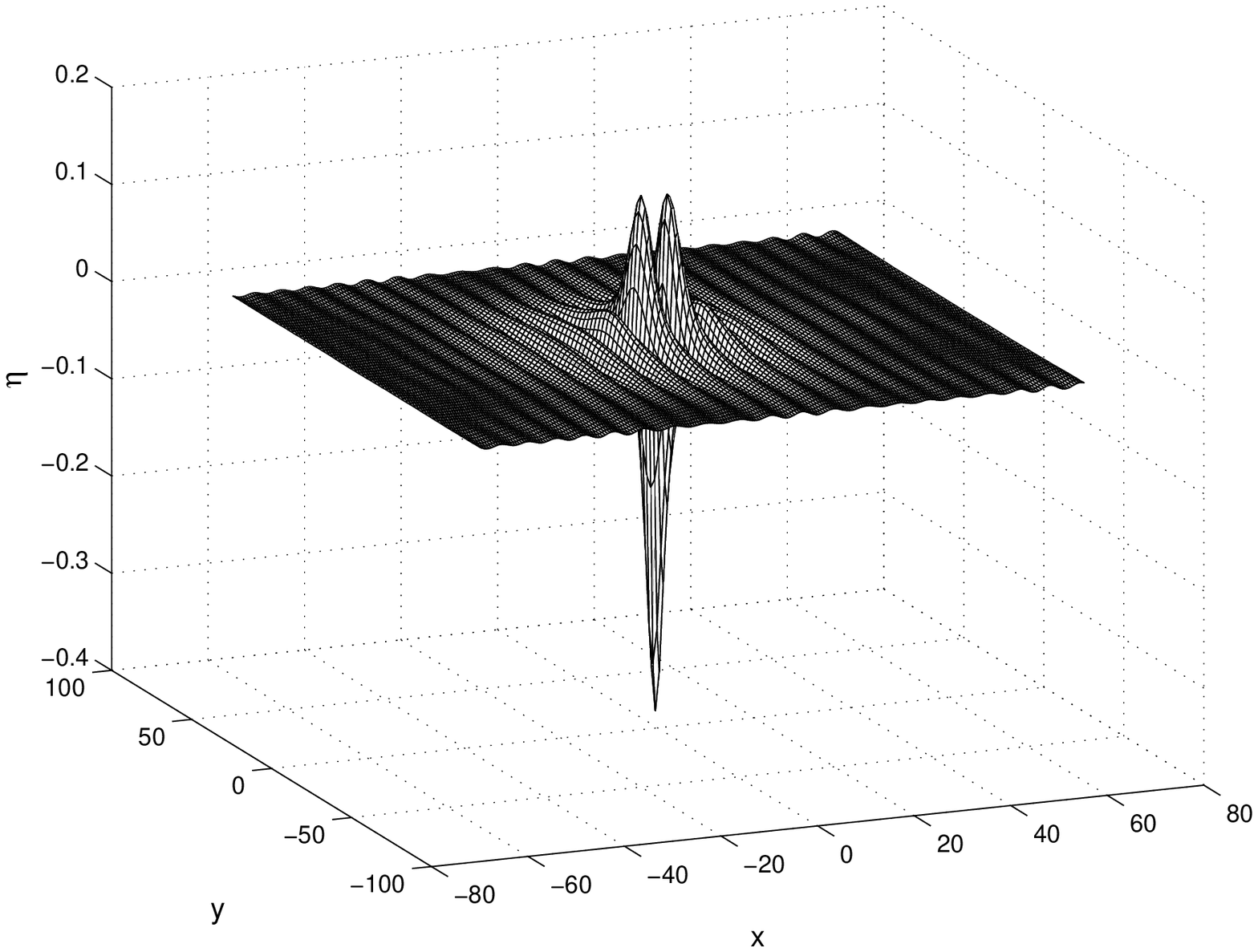,width=0.48\textwidth}

\psfig{file=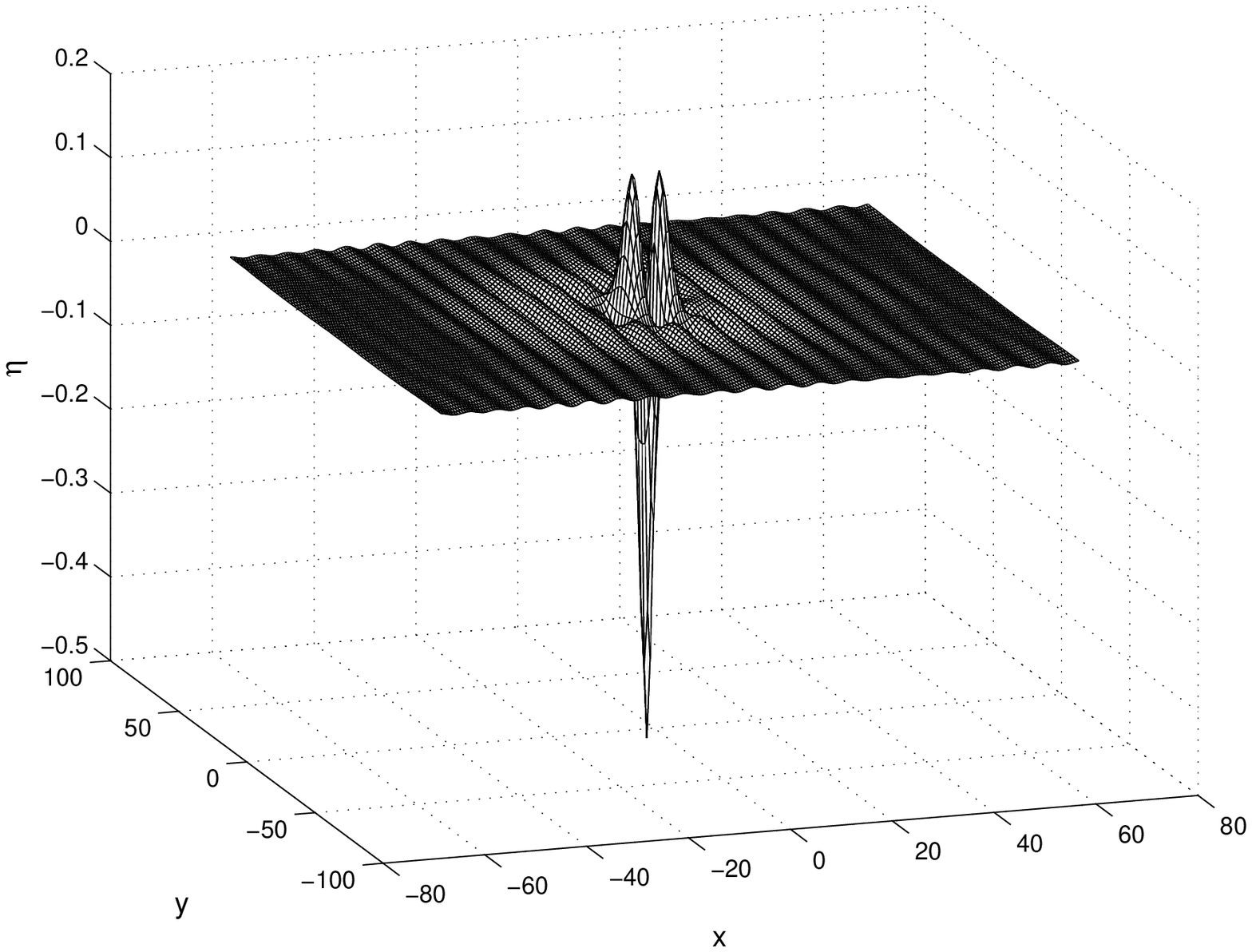,width=0.48\textwidth} \quad
\psfig{file=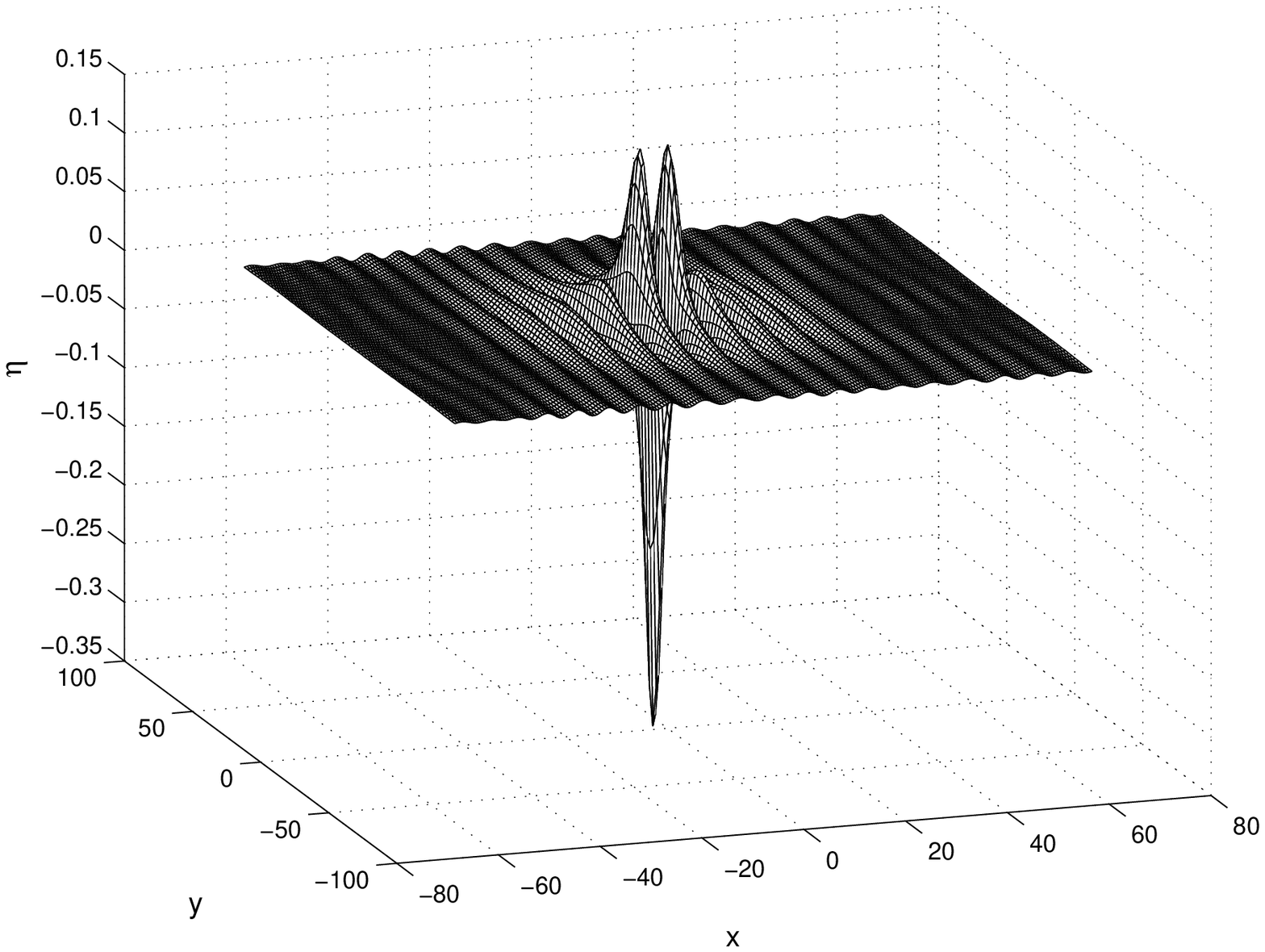,width=0.48\textwidth}
\end{center}
\caption{From top-left to bottom-right: snapshots of one period of a breather taken at $t=1240,1440,1520,1620$ for the small amplitude negative energy computation discussed in Figure \ref{DepLumpStab}.}
\label{figBreather}
\end{figure}

The stability of localised traveling waves is considered next.
First, the virial argument sketched above implies that near the
bifurcation point, both elevation and depression solitary waves are
linearly unstable. The virial argument would predict eventual blowup for
negative energy perturbations and this is not observed, in all cases the blowup 
is arrested by the generation of a larger amplitude breather. Dispersive spreading 
consistent with the envelope spreading predicted by the virial argument is observed for positive energy perturbations. 
In the computations that we present we perturb the exact solitary wave
solution by a small multiple of itself, by taking initial data
$\eta(x,y,0) = (1+\delta)\widetilde{\eta}$ where $\widetilde{\eta}$
is the computed travelling wave. Since, for the perturbed wave,
$$E= \int|\nabla A|^2 - \frac12 |A|^4 \approx  - \delta \int  |\widetilde{A}|^4,$$
where $\widetilde{A}$ is the envelope of the solitary wave, negative energy perturbations
correspond to $\delta>0$ and positive energy ones to $\delta<0$.
In the left panel of Figure \ref{DepLumpStab} we show the typical evolution of the 
amplitude for a perturbed small amplitude depression solitary wave. For a negative 
energy perturbation we see focussing arrested by the formation of a breather whereas for 
positive energy the amplitude decreases monotonically as a result of dispersive spreading. The case shown is for a 
depression wave, however the elevation wave dynamics is broadly similar. In Figure \ref{figBreather} we show four snapshots of the
breather evolution resulting from the unstable small amplitude depression wave at later times.

For a large class of problems, linear
stability may change at a critical point of $H(c)$, the speed-energy
curve. This necessary condition for instability was observed
by \cite{S}, for gravity waves and has since been extended to many other
situations. {\cite{AC}} present an application of this result to a wavepacket solitary wave
in a model equation. In the present problem, both depression and
elevation waves have critical points in $H(c)$. We denote solitary
waves with speed lower than this critical speed ``large amplitude"
and those with speed larger than it ``small amplitude" since in all our computations 
amplitude is a monotonic function of speed. Small amplitude waves' bifurcation diagram and linear stability are well described by the NLS equation whereas large amplitude waves' are not.
Our numerical simulations show that for depression waves, the
exchange of stability does take place at the minimum point in
$H(c)$. The evolution of large amplitude depression waves'
amplitudes when subject to perturbations is shown in the right panel of Figure
\ref{DepLumpStab}. The waves are stable regardless of the sign of
$\delta$. Solitary waves (or breathers which are small perturbations of the solitary waves) 
propagate in the midst of small linear dispersive waves that have been shed by the perturbed initial data. 

\begin{figure}
\begin{center}
\psfig{file=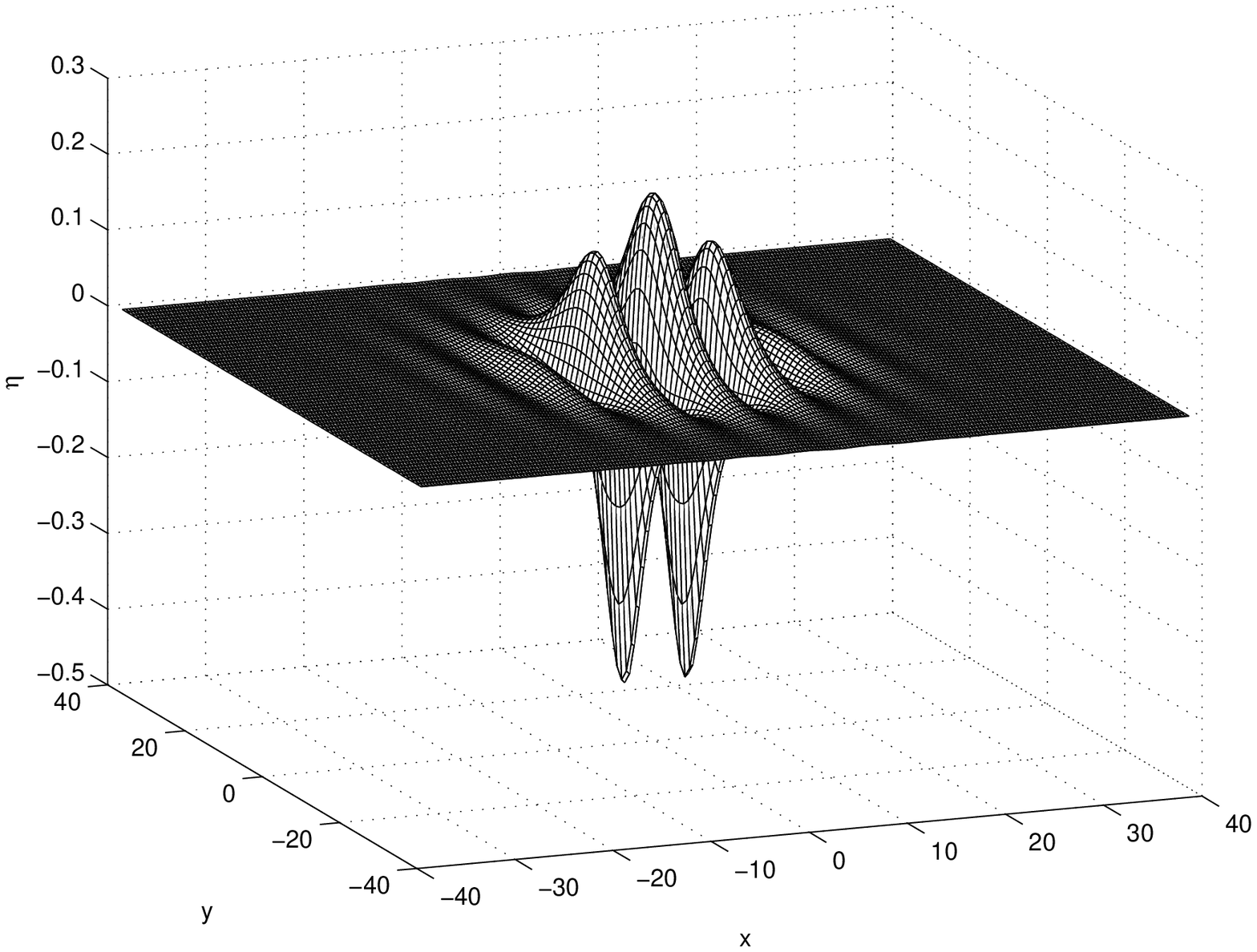,width=0.48\textwidth} \quad
\psfig{file=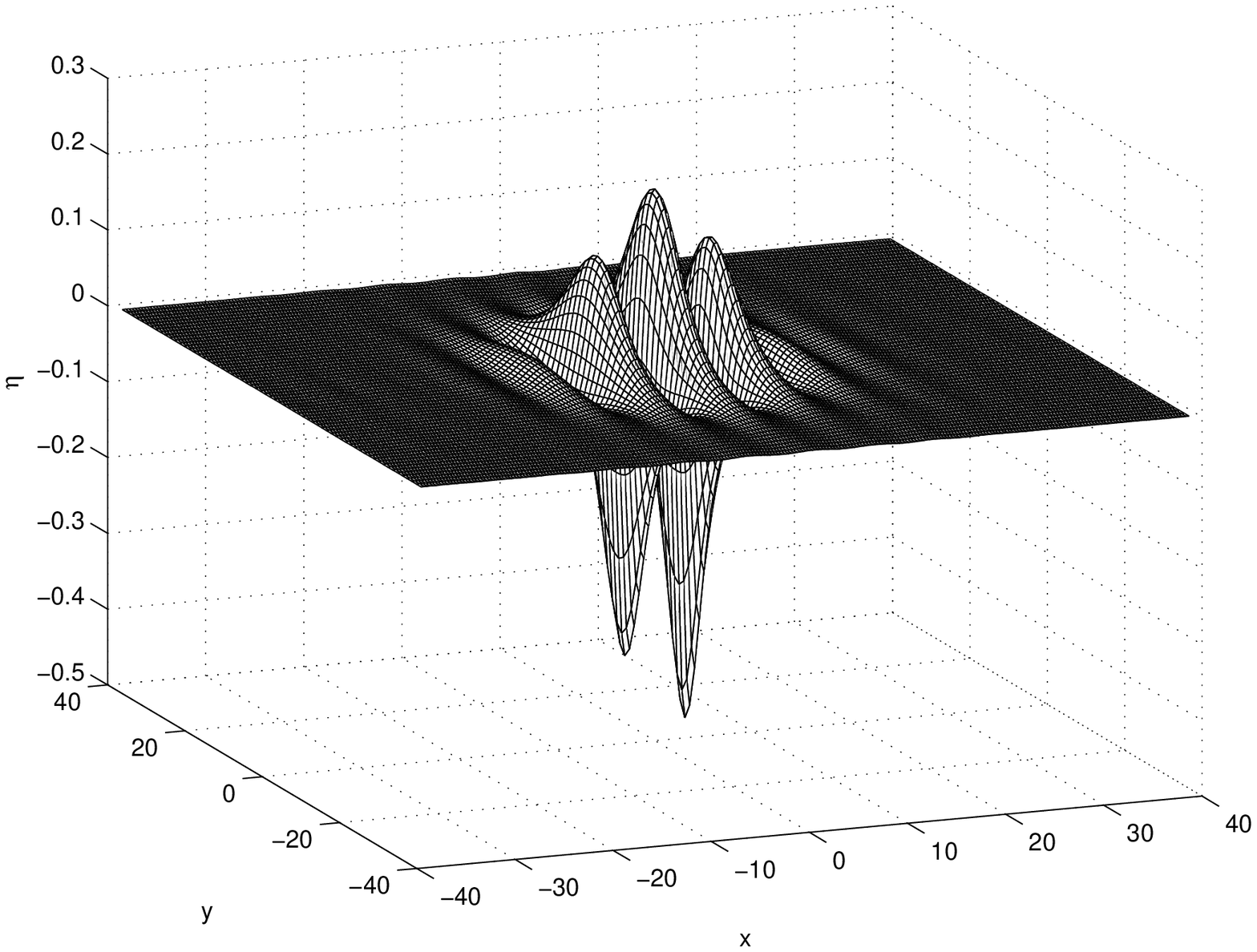,width=0.48\textwidth}

\psfig{file=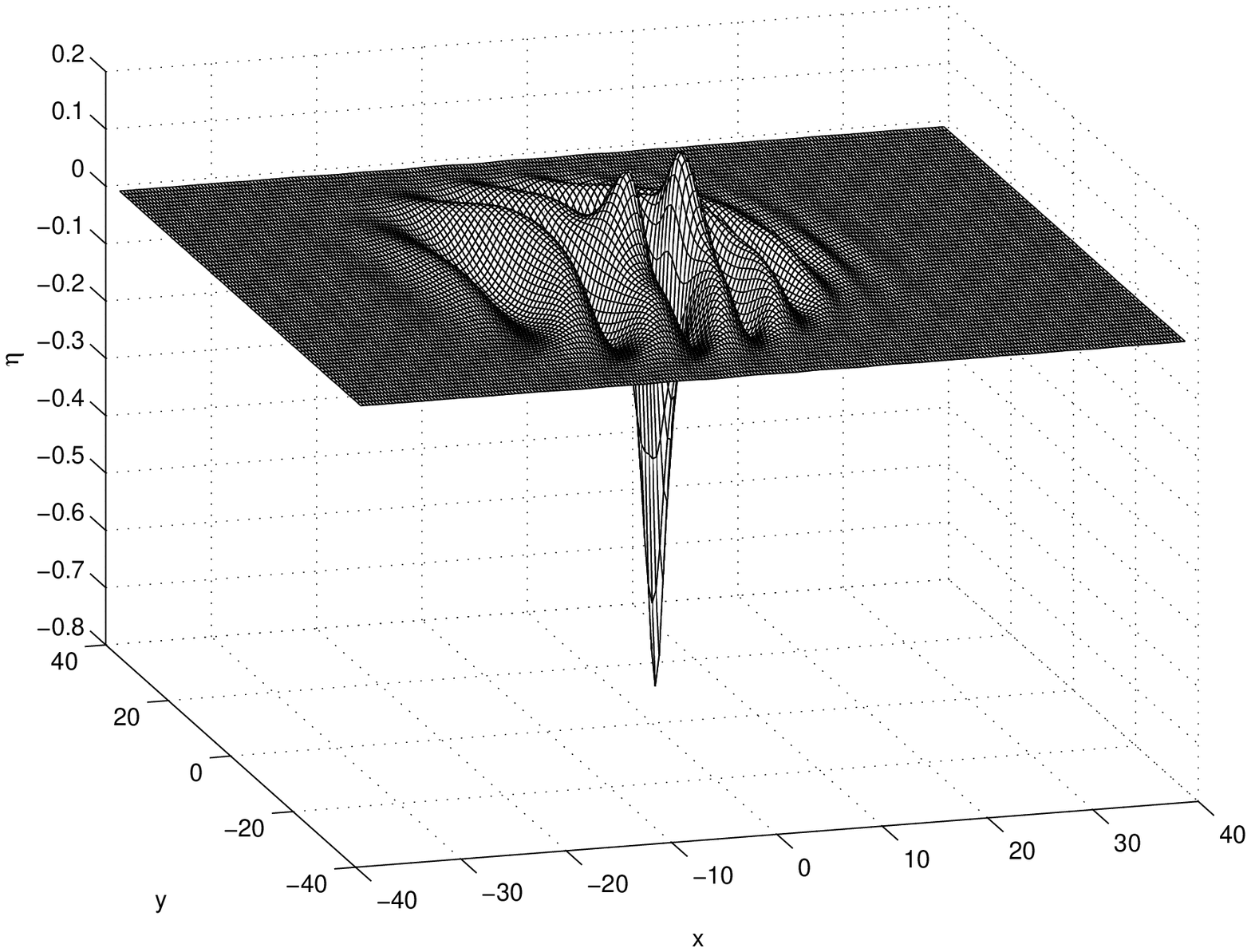,width=0.48\textwidth} \quad
\psfig{file=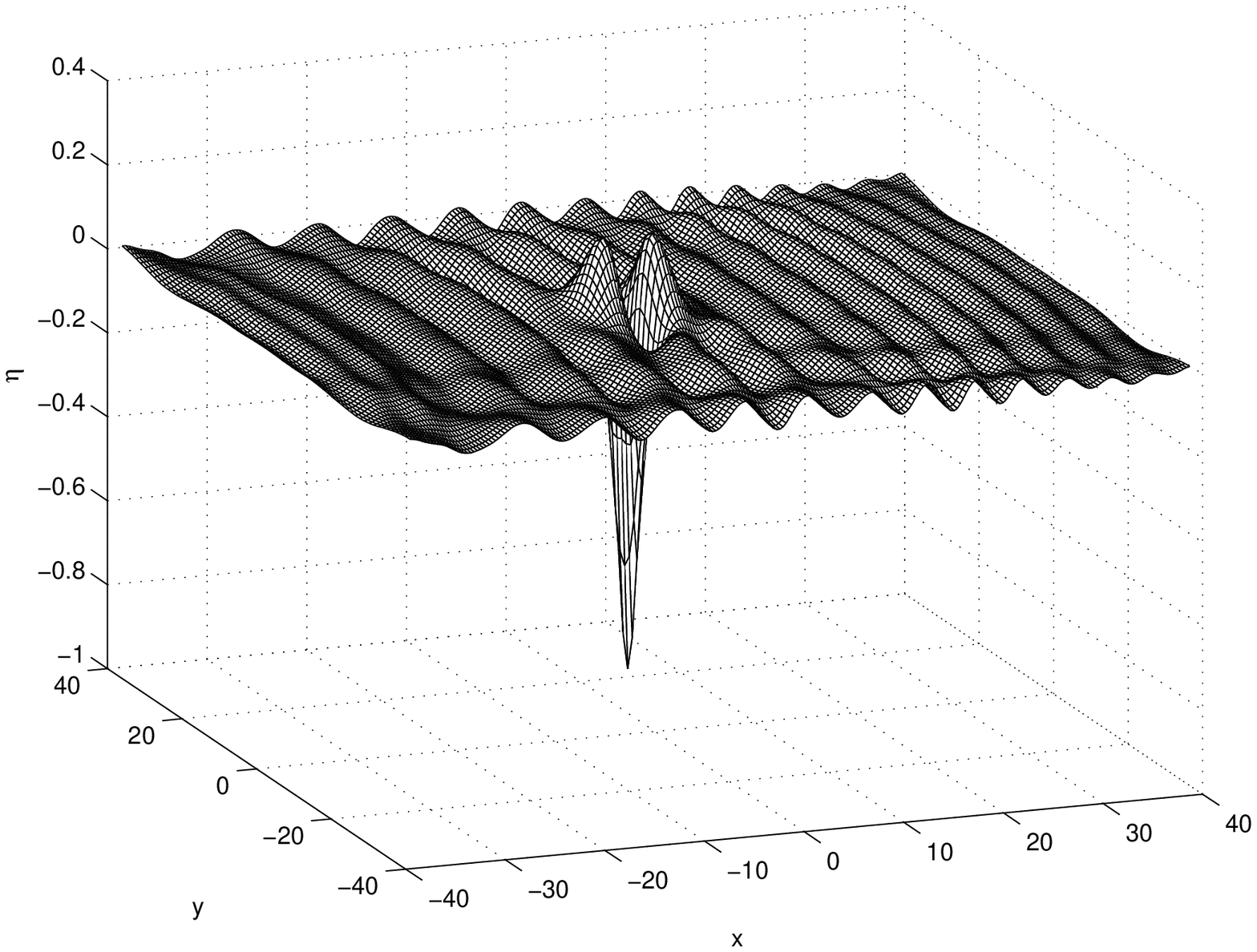,width=0.48\textwidth}
\end{center}
\caption{From top-left to bottom-right: snapshots of the evolution of a large amplitude unstable
elevation wave with $\eta(0,0)=0.2196$, $c=1.3907$ in cDtNE at $t=0,120,200,270$. The perturbation is 0.01 times the original solitary wave. Some waves are radiated during the transition and the unstable wave evolves into a breather.}
\label{figUnEleProf}
\end{figure}

All our computations show that elevation waves remain unstable at large amplitude.
In Figure \ref{figUnEleProf} an unstable large elevation solitary wave subject to a smallnegative energy perturbation
and evolves into a depression breather. Similarly, positive energy perturbations will also yield depression breathers (whose energy is much smaller than the elevation waves (see Figure \ref{figBif2D}). The instability is initially similar to that of a 1D depression wave (see \cite{MVW}): a symmetry breaking whereby the leading trough grows at the expense of
the trailing one (this instability is associated to the translational invariance mentioned above). What follows is
collapse focussing that is arrested by the formation of a depression breather.  

From our calculations, we believe there are families of periodic (breather) solutions of different
periods and amplitude for each fixed energy above the minimum of the depression solitary waves. 
The orbits of these breathers in phase space are centred around the fixed point of stable depression solitary waves and the precise bifurcation diagram for them would require computing 
exact periodic localised structures which is beyond the scope of this paper. 
A summary of the stability results in this section is shown in table \ref{table}.

\begin{table}
\begin{center}
\begin{tabular}{@{}lcc@{}}
&positive energy perturbation & negative energy perturbation\\ [3pt]
small depression & unstable$\rightarrow$large breather &unstable$\rightarrow$disperses out\\
small elevation  & unstable$\rightarrow$large breather &unstable$\rightarrow$disperses out\\
large depression & stable & stable\\
large elevation  & unstable$\rightarrow$large breather &unstable$\rightarrow$large breather\\
\label{table}
\end{tabular}
\end{center}
\caption{Summary of stability Results}
\end{table}

\begin{figure}
\begin{center}
\psfig{file=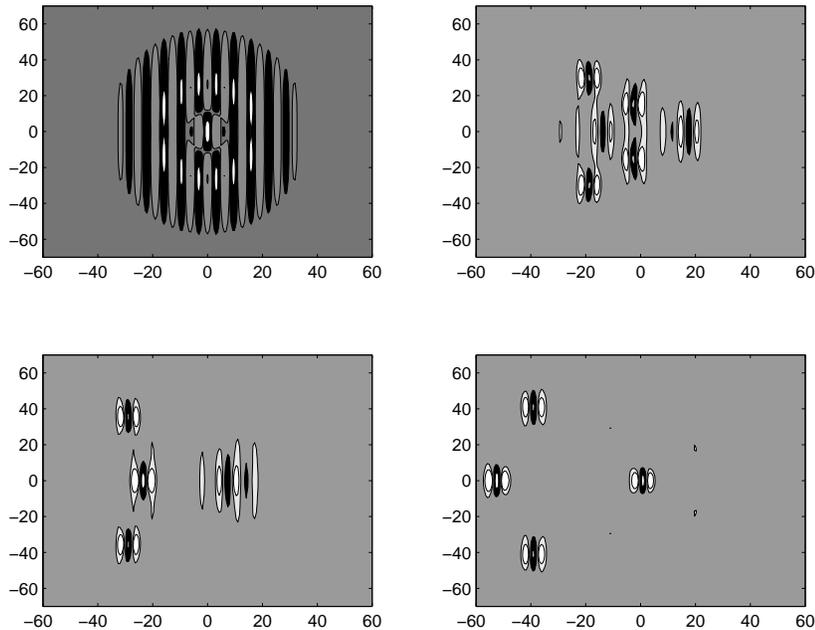,width=0.8\textwidth}\end{center}
\caption{Instability of a higher energy solitary wave with $\eta(0,0)=-0.4028, c=1.4061$. Snapshots of the evolution are shown at $t=0$ (top left), $t=800$ (top right), $t=1200$ (bottom left), $t=1600$ (bottom right). Four depression breathers emerge from the complex behaviour.} \label{FigMode2}
\end{figure}

Lastly we compute the evolution of a complex solitary wave of the
cDtNE equations corresponding to a higher mode of the nonlinear
eigenvalue problem discussed previously, and this is shown in
Figure \ref{FigMode2}. In this case all perturbations triggered
rapid instabilities, but the nonlinear evolution shows a remarkable
dynamics with eventual focussing into four depression
breathers of different amplitudes.

\subsection{Collisions}

\begin{figure}
\begin{center}
\psfig{file=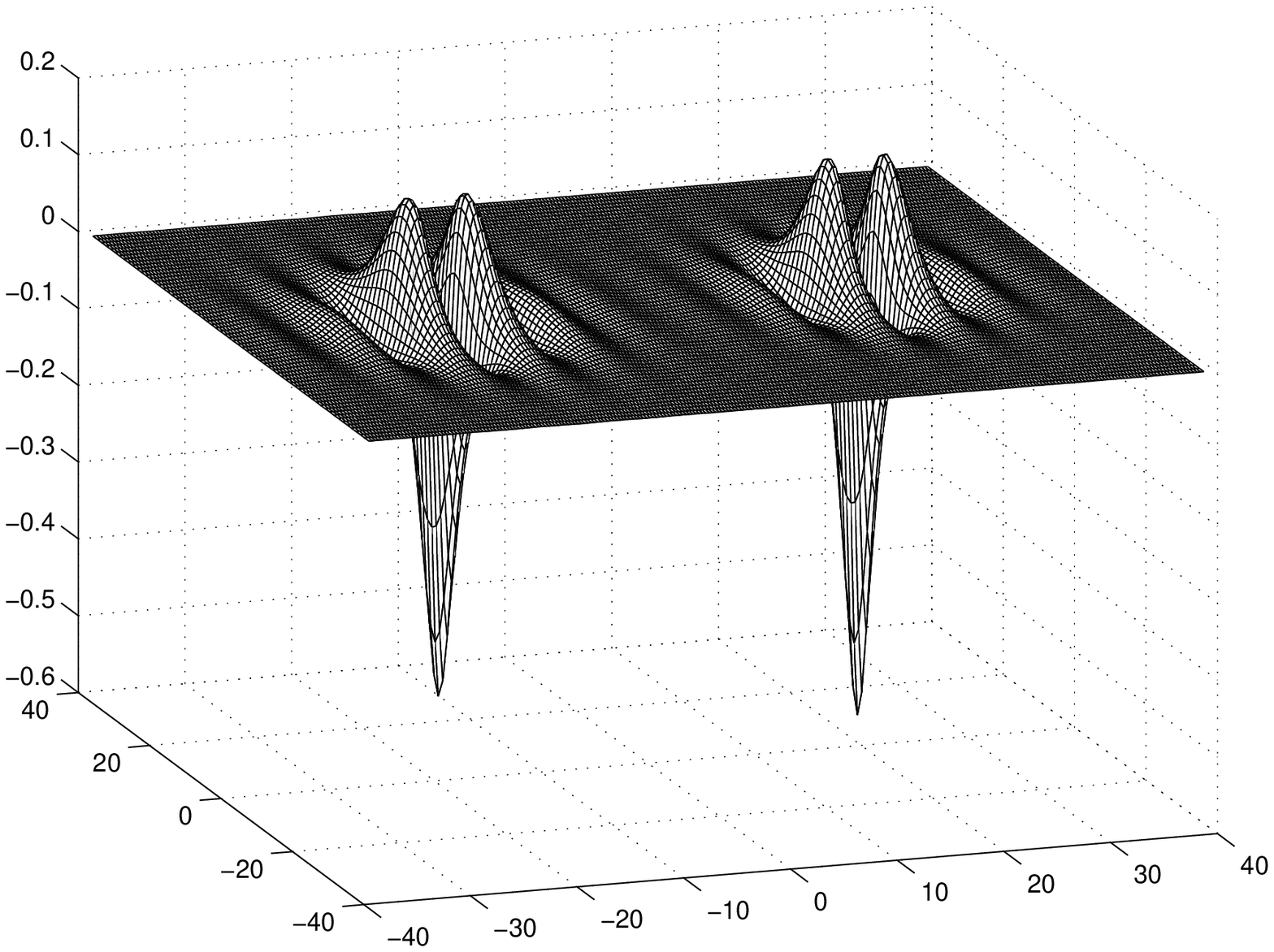,width=0.48\textwidth} \quad
\psfig{file=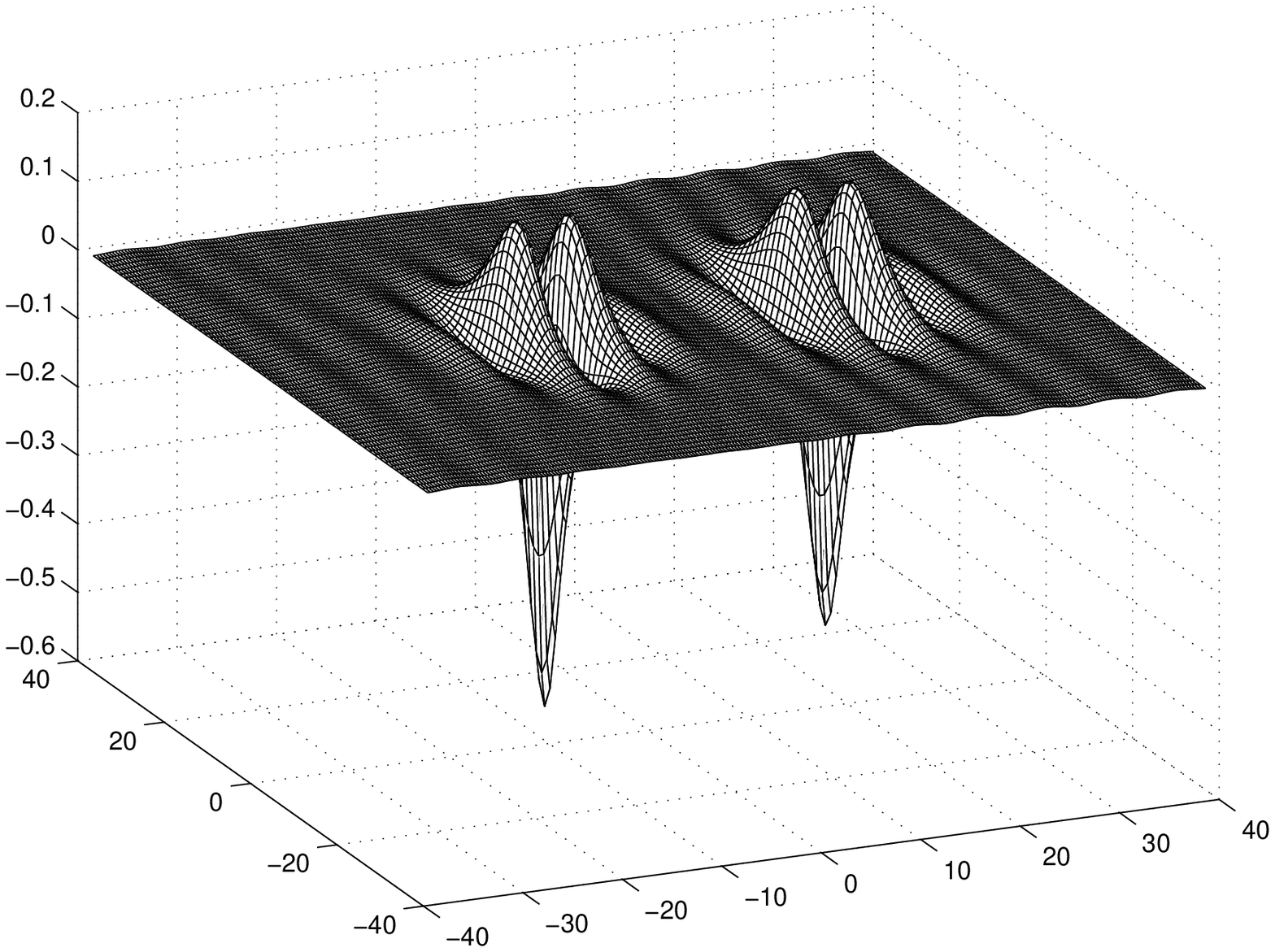,width=0.48\textwidth}
\end{center}
\caption{Collision of two stable depression solitary waves travelling in the positive x-direction. Left: Initial data consisting of a shifted superposition of two waves with with $\eta(0,0)=-0.56, c=1.3782$ and $\eta(0,0)=-0.49, c=1.3873$. Right: Solution after interaction at $t=4000$. Note that the collision is not completely elastic, with their amplitudes decreasing slightly as a result of the collision. The solution was computed in a frame of reference moving at speed $1.3828 $.} \label{colldd}
\end{figure}

\begin{figure}
\begin{center}
\psfig{file=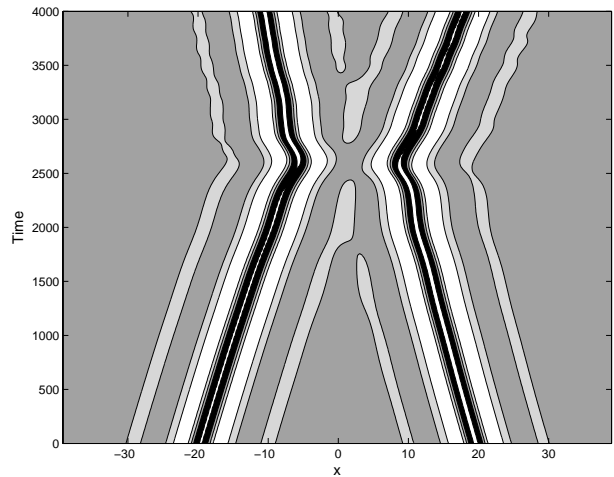,width=0.6\textwidth}
\end{center}
\caption{Filled contour plot of the centreline of the dynamics $\eta(x,0,t)$ for the collision reported in Figure \ref{colldd}. Note that the waves' core remain distant during the collision and that after the collision the waves have exchanged identities and undergone a phase shift.} \label{collddc}
\end{figure}

Given the stability of large amplitude depression waves, we have
numerically computed their collisions. We have only computed head-on
and overtaking collisions of pairs of waves although in a 2D problem
there is a wide range of possible collision scenarios. For head-on
collisions, the interaction time between the two waves is
insufficient for any strong nonlinear effect to take place and we
have only observed very small oscillations that result from the
small inelasticity of the collision. The waves essentially traverse each
other. The more interesting case is the overtaking collision. Here,
the small difference in solitary wave speeds implies that the
calculations must be carried out over long times. In the 1D problem
collisions were of two types of inelastic collisions \cite{MVW}: collisions where
both waves survived and collisions where only the larger wave survived when
their amplitude difference was large. Here, we have only observed
quasi-elastic collisions where both waves survive and the primary effect of the
overtaking collision is a rapid and large phase shift of the order
of one envelope wavelength. Figure \ref{colldd} shows the before and
after free surface profiles of the waves and Figure \ref{collddc}
shows the resulting waves' trajectories in $(x,t)$ space. The
collisions are weakly inelastic and the wave amplitude is mildly attenuated
with some dispersive radiation present during the collision.

\section{Extensions}\label{extensions}

There are several extensions of the model which, while the detailed
study is beyond the scope of this paper, may be useful in studying
problems of this type. We have considered fluids of infinite depth
only (which in the case of GC flows in water is a good
approximation for depths exceeding a few centimetres). If the effect
of finite depth is required, the formulation can be modified simply
by taking
$G_0(\eta)=|D|\tanh({h}|D|)$ where
${h}$ is the mean depth of the fluid
normalised by the capillary gravity length scale and is inversely
related to the Bond number. In particular the shallow water lump
dynamics modelled by the Kadomtsev-Petviashvilli (KP) equation should be
recovered when the {Bond} number is
greater than $1/3$.

Given the small length scales of the waves, it may also be of
practical interest to add the effects of viscosity and forcing. In the analysis of their experiments \cite{CDAD2} adjusted a simple one-way model proposed in \cite{AM2} by adding forcing and a viscous damping. The more accurate cDtNE equations can be modified to include
small viscous damping effects by using the approximation presented in {\cite{DDZ}} whereby dissipation is modelled through the modification of both
kinematic and dynamic boundary conditions:
\begin{eqnarray}
\eta_t - G_0 \xi &=& 2\Rey^{-1} \Delta\eta + (G_1+G_2)\xi \\
\xi_t +  (1 - \Delta) \eta &=& 2\Rey^{-1} \Delta\xi + P(x,y,t) + \nabla\cdot\Big[\frac{\nabla\eta}{\sqrt{1+|\nabla\eta|^2}} - \nabla \eta \Big]\nonumber\\
&&+\frac{1}{2}\Big[(G_0\xi)\big(G_0\xi-2G_0\eta
G_0\xi-2\eta\Delta\xi\big)-|\nabla\xi|^2\Big].
\end{eqnarray}
Here, $P(x,y,t)$ is the pressure forcing and the Reynolds number, which controls the dissipation rate, is given by 
$$\Rey = \frac{VL}{\nu} = \frac{(\sigma/\rho)^{3/4}}{g^{1/4} \nu},$$
where $\nu$ is the kinematic viscosity of the fluid. For the case of an air-water interface in the regime that we considered in this paper, we obtain a Reynolds number of approximately 500.

For small amplitude waves, the variation in the transverse direction is not significant compared to that in the propagation direction. One can therefore propose to assume this a priori and simplify the system by deleting all the nonlinear terms which include y-derivatives obtaining a ``weakly transversal" model which still has the same NLS equation describing wave packets as the full problem. The Hamiltonian for this model reads
\begin{eqnarray}
\overline{H}[\eta,\xi]=\int\:\frac{1}{2}\xi\big(G_0+\overline{G}_1
+\overline{G}_2\big)\xi+\frac{1}{2}\eta^2+ \frac12\eta_y^2 +\big(\sqrt{1+\eta_x^2}\big)\:dxdy
\end{eqnarray}
where
\begin{eqnarray}
\overline{G}_0&=&\big(-\p_{xx}\big)^{1/2}\nonumber\\
\overline{G}_1&=&-\p_x\eta\p_x-\overline{G}_0\eta\overline{G}_0\nonumber\\
\overline{G}_2&=&\frac{1}{2}\overline{G}_0\eta^2\p_{xx}
+\frac{1}{2}\p_{xx}\eta^2\overline{G}_0+\overline{G}_0\eta\overline{G}_0\eta\overline{G}_0
\end{eqnarray}
Calculations performed with this approximation show excellent agreement with cDtNE at small amplitudes and only qualitative agreement for larger amplitude solitary waves.

\section{Conclusions}\label{conclusions}
The dynamics of Gravity-Capillary solitary waves on the surface of
three-dimensional fluid was studied. The only approximation made was
a cubic truncation of a scaled  Dirichelet to Neumann map that
provides the normal velocity of the free-surface given its
tangential velocity. In two dimensions, where comparisons to the
untruncated problem (i.e. fully nonlinear free-surface potential
flow) can be accurately measured, this truncation is remarkably
accurate in modelling small and moderate amplitude waves. We
conjecture that the same is true in three-dimensions.

There are undoubtedly in this problem infinitely many branches of
solitary travelling wave solutions as this is the prediction of the
associated envelope NLS analysis about the bifurcation point. We
compute examples of three of them: elevation and depression solitary
waves arising from the simplest eigenfunction of the NLS problem and
a more complex wave arising from a higher eigenfunction of the same
problem. The localised solitary waves have the surprising property
that as their amplitude (from peak to trough) decreases to zero as
the bifurcation point is approached, their physical energy tends to
a finite positive value, quantised by the different eigenfunctions
of the NLS equation.

The instability and subsequent evolution for one dimensional line
solitary waves and the various two dimensional solitary waves have
been explored numerically by perturbing the waves and computing the
solution through accurate pseudo spectral based methods. All
solitary waves are found to be unstable with the notable exception
of larger amplitude depression waves. These waves together with
travelling breathers, which are periodic travelling cycles
oscillating about these travelling states, are stable and appear to
be attractors in the long time evolution of the problem.

The focussing NLS equation adequately predicts the bifurcation and linear stability properties of small amplitude solitary CG waves. The collapse singularity of initial data, however, is not observed in our computations leading to the conjecture that an appropriate envelope model for this problem is a cubic-quintic NLS equation where the quintic term is defocussing.

\smallskip
{\bf Acknowledgements}
\smallskip

We thank Dr E. P$\breve{a}$r$\breve{a}$u for making available more resolved numerical results solitary waves. This work was supported by EPSRC, under Grant Number GR/S47786/01, by the Division of Mathematical Sciences of the National Science Foundation, under Grant Number DMS-0908077, and by a Royal Society Wolfson award.

\bibliographystyle{abbrv}

\end{document}